\documentclass[sigconf]{acmart}
\AtBeginDocument{%
  }

\copyrightyear{2025}
\acmYear{2025}
\setcopyright{acmlicensed}\acmConference[KDD '25]{Proceedings of the 31st ACM
SIGKDD Conference on Knowledge Discovery and Data Mining V.1}{August 3--7,
2025}{Toronto, ON, Canada}
\acmBooktitle{Proceedings of the 31st ACM SIGKDD Conference on Knowledge
Discovery and Data Mining V.1 (KDD '25), August 3--7, 2025, Toronto, ON, Canada}
\acmDOI{10.1145/3690624.3709172}
\acmISBN{979-8-4007-1245-6/25/08}

\usepackage{enumitem}
\usepackage{hyperref}
\usepackage{multirow}
\usepackage{booktabs}
\usepackage{listings}
\usepackage{xcolor}

\usepackage{amsthm}

\usepackage{algorithm}
\usepackage{algorithmic}

\usepackage{listings}

\begin{document}

\title[Denoising Programming Knowledge Tracing with a Code Graph-based Tuning Adaptor]{Denoising Programming Knowledge Tracing\\with a Code Graph-based Tuning Adaptor}

\author{Weibo Gao}
\orcid{0000-0003-0894-7023}
\affiliation{%
  \institution{State Key Laboratory of Cognitive Intelligence, University of Science and Technology of China}
  \city{Hefei}
  \country{China}
}
\email{weibogao@mail.ustc.edu.cn}

\author{Qi Liu}
\authornote{Corresponding Author.}
\orcid{0000-0001-6956-5550}
\affiliation{%
  \institution{State Key Laboratory of Cognitive Intelligence, University of Science and Technology of China}
  \institution{Institute of Artificial Intelligence, Hefei Comprehensive National Science Center}
  \city{Hefei}
  \country{China}}
\email{qiliuql@ustc.edu.cn}

\author{Rui Li}
\orcid{0009-0005-3657-1133}
\affiliation{%
  \institution{State Key Laboratory of Cognitive Intelligence, University of Science and Technology of China}
  \city{Hefei}
  \country{China}
}
\email{ruili2000@mail.ustc.edu.cn}

\author{Yuze Zhao}
\orcid{0009-0007-3542-4304}
\affiliation{%
  \institution{State Key Laboratory of Cognitive Intelligence, University of Science and Technology of China}
  \city{Hefei}
  \country{China}
}
\email{yuzezhao@mail.ustc.edu.cn}

\author{Hao Wang}
\orcid{0000-0001-9921-2078}
\affiliation{%
  \institution{State Key Laboratory of Cognitive Intelligence, University of Science and Technology of China}
  \city{Hefei}
  \country{China}
}
\email{wanghao3@ustc.edu.cn}

\author{Linan Yue}
\orcid{0000-0002-5980-6098}
\affiliation{%
  \institution{State Key Laboratory of Cognitive Intelligence, University of Science and Technology of China}
  \city{Hefei}
  \country{China}
}
\email{lnyue@mail.ustc.edu.cn}

\author{Fangzhou Yao}
\orcid{0000-0002-5085-7841}
\affiliation{%
  \institution{State Key Laboratory of Cognitive Intelligence, University of Science and Technology of China}
  \city{Hefei}
  \country{China}
}
\email{fangzhouyao@mail.ustc.edu.cn}

\author{Zheng Zhang}
\orcid{0009-0002-8689-0763}
\affiliation{%
  \institution{State Key Laboratory of Cognitive Intelligence, University of Science and Technology of China}
  \city{Hefei}
  \country{China}
}
\email{zhangzheng@mail.ustc.edu.cn}

\renewcommand{\shortauthors}{Weibo Gao et al.}

\begin{abstract}
Programming Knowledge Tracking (PKT) aims to dynamically diagnose learners' mastery levels of programming knowledge based on their coding activities, facilitating more effective and personalized programming education. However, current PKT studies primarily focus on the implicit relationship between code content and knowledge assessment, often overlooking two types of noise signals in long-term programming activities: unwanted signals from unrelated submissions and weak signals from minor modifications. This practical challenge significantly limits model performance and application.
To address this issue, we propose \textbf{Coda}, a \textbf{Cod}e graph-based tuning \textbf{a}daptor designed to enhance existing PKT models by identifying and mitigating the impact of noise. Specifically, Coda first transforms the loose code sequences submitted by each learner into a compact code graph. By leveraging this code graph, unwanted signals can be identified from a semantic similarity perspective. We then apply a cluster-aware GCN to the code graph, which improves the discrimination of weak signals and enables their clustering for identification. Finally, a lightweight yet effective adaptor is incorporated into the PKT task through optimization with two noise feature-based constraints and a navigational regularization term, to correct knowledge states affected by noise.
It is worth mentioning that the Coda framework is model-agnostic and can be adapted to most existing PKT solutions. 
Extensive experimental results on four real-world datasets demonstrate that Coda effectively performs the PKT task in the presence of noisy programming records, outperforming typical baselines.
\end{abstract}




\maketitle

\section{Introduction}
\label{sec:intro}

With the advent of online programming platforms like $LeetCode.com$, it has become increasingly popular for students and other learners to use these websites for coding training~\cite{yin2023tracing, han2023errorclr,Liu2024Icdm}. As data on online programming activities accumulate, understanding the programming training process is crucial for enhancing online learning services, such as question recommendation~\cite{wu2020exercise, zhao2023cross}.

\begin{figure}
    \centering
    \scalebox{0.29}
    {\includegraphics{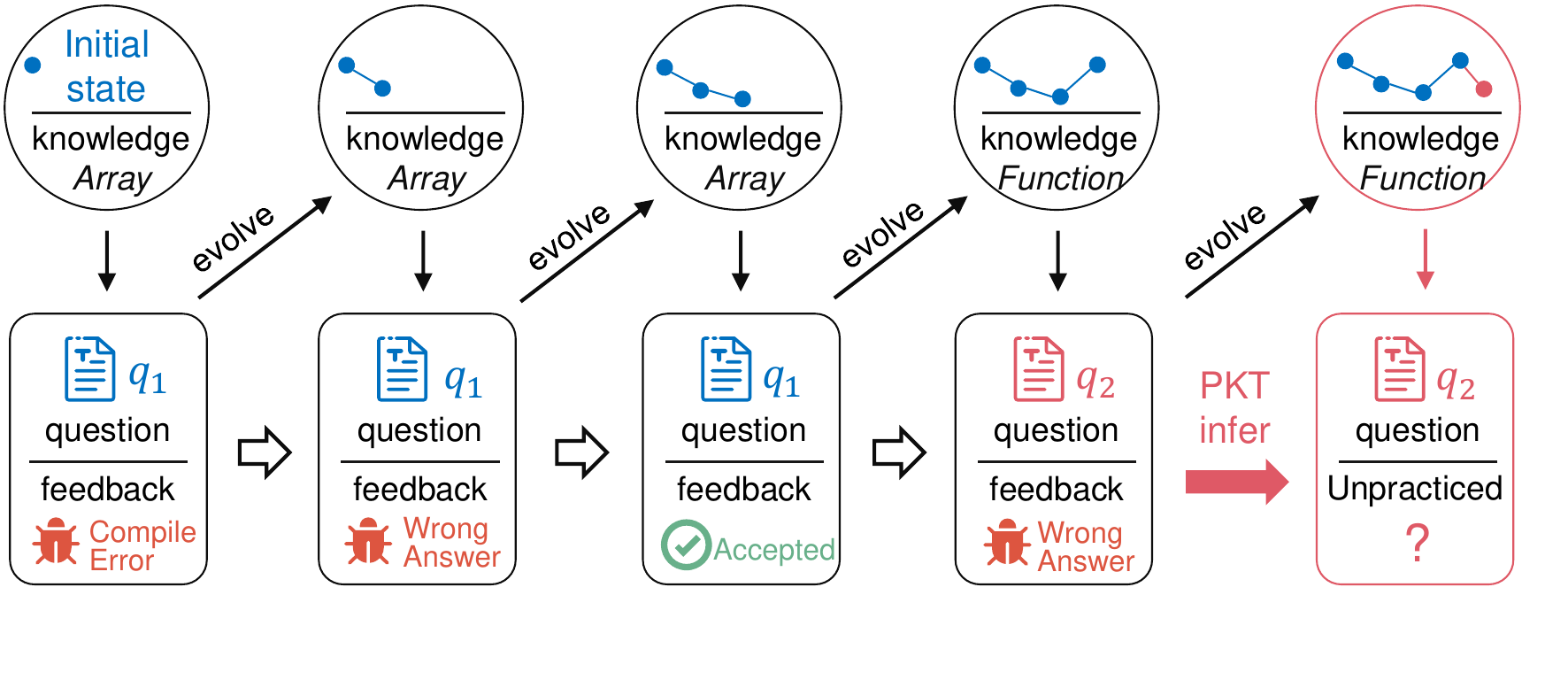}}
    \vspace{-1cm}
    \caption{The example of the programming practice process, where a PKT model is used to infer future performance.}
    \label{fig:intro}
    \vspace{-0.5cm}
\end{figure}

Human programming practice generally involves an iterative process of code submission and modification to solve tasks. Learners submit code and receive feedback, such as error messages like \textit{Compile Error}, or positive feedback like \textit{Accepted}. Based on the specific feedback, they revise their previous submissions. Figure~\ref{fig:intro} illustrates this process, where a learner attempts to solve each question (e.g., $q_{1}$) through multiple submissions. During this process, learners develop an internal understanding of the question, or technically speaking, a hidden computational knowledge state, representing their learning progress. As learners refine their code based on feedback, their knowledge states evolve, reflecting their learning journey. To better understand programming learning, it is essential to model and trace the evolution of these invisible knowledge states, which is the focus of Programming Knowledge Tracing\cite{zhu2022programming}.

Programming Knowledge Tracing (PKT) involves dynamically evaluating learners' programming knowledge states based on their coding activities. These knowledge states refer to learners' proficiency in various programming knowledge concepts such as \textit{Array} and \textit{Function}. PKT extends traditional knowledge tracing (KT) techniques~\cite{shen2024survey,wang2022neuralcd}, which are designed for non-programming general-purpose learning environments~\cite{li2022pst}, by incorporating code-related features specific to programming environments. Its tracing results offer significant insights into students' coding abilities~\cite{lei2021systematic, zhu2022programming, liu2022gpt}. Since knowledge states cannot be directly observed, most attempts model PKT as a next-performance prediction problem, usually using recurrent neural networks (RNNs) to summarize programming sequences~\cite{yin2023tracing}. These models mine code features to infer learner knowledge states from their codes\cite{li2022pst,liang2022help,shi2022code}, establishing implicit relationships between code and knowledge~\cite{zhu2022programming}, and achieving high predictive performance. However, these models are limited by the numerous noisy submissions in real-world programming data, which stem from the open-ended nature of programming platforms.

Through examining online programming environments~\cite{nguyen2014codewebs, zinovieva2021use}, we identify two primary categories of prevalent noisy signals during programming:
\vspace{-0.5cm}
\begin{itemize}[leftmargin=10pt]
    \item \textbf{Unwanted signals from unrelated submissions.}
    In the majority of learners' programming practice records, there are commonly codes that are unrelated to the solved questions. Typical examples include:
    (1) Novice programmers are unfamiliar with the web-based coding environment and use simple but irrelevant code snippets (e.g., ``\textit{print (`Welcome to KDD 2025')}'') to test the use of the online coding environment.
    (2) Learners may carelessly submit codes to inappropriate questions. For instance, a learner might mistakenly submit the solution for the question ``\textit{Add Two Numbers}'' to a similar question ``\textit{Add Binary}'' when both submission pages are open simultaneously in their web browser.
    These records do not reflect the learner's knowledge and introduce unwanted signals in their programming history, complicating the accurate modeling of true knowledge states.
    \item \textbf{Weak signals from minor modifications.}
    Solving a programming problem typically involves multiple submissions, where many submissions are minor modifications of previous ones based on feedback. After several failed attempts, learners may submit an entirely new code. We argue that learners' knowledge states are likely to remain unchanged for minor modification submissions, as the modifications are not significant improvements. While these minor changes may be correct or meaningful for humans, they are redundant to PKT models.
    These weak signals can mislead the PKT model if used directly in training, leading to overestimation or underestimation of the learner's knowledge state.
\end{itemize}

\textbf{Note:} The noisy signals described above provide two typical insights for our preliminary exploration. Many other types of noise exist in programming data, which we plan to explore in future research. Additionally, although these so-called noisy code contents may be meaningful from a human perspective, they could still represent irrelevant or redundant signals for current PKT technology, so we classify them as noise.

In this paper, our goal is to mitigate the impact of these noisy signals to enhance existing PKT models, which presents two primary challenges:
(1) Accurately identifying noise signals is challenging due to the absence of noise labels or noise annotation criteria. The available clues are the submitted codes, feedback, and question information.
(2) Designing an effective and lightweight method to use noise identification to denoise existing PKT solutions is significant, due to the computational complexity of PKT tasks.

To address these challenges, we propose \textbf{Coda}, a \textbf{cod}e graph-based tuning \textbf{a}daptor designed to enhance existing PKT models through denoising. While mature denoising methods exist in other fields~\cite{ho2020denoising}, this is the first attempt to address denoising within PKT. Our model comprises two key stages: the \textit{noisy signal identification stage} and the \textit{adaptive PKT stage}. In the noisy signal identification stage, we construct a code graph based on programming sequences to explore semantic relations between codes. Using this code graph, we assess unwanted signals from unrelated codes by examining semantic similarities between codes and their correlations with questions. To distinguish weak signals from submissions, we employ a cluster-aware Graph Convolution Network (GCN) over the code graph, which improves the discrimination of weak signals and enables their clustering for identification. 
In the adaptive PKT stage, a tuning adaptor facilitates the adaptation of the PKT task to the noisy signals by tuning a trained PKT backbone model, ensuring computational efficiency. Two noise code feature-based constraints and a navigational regularization term with theoretical derivation guide the tuning process, ensuring effective adaptation of PKT.
It is worth mentioning that the Coda framework is model-agnostic and can be adapted to most existing PKT models. Extensive experiments on four real-world datasets demonstrate that Coda effectively performs the PKT task in the presence of noisy programming records, surpassing typical baselines.
\section{Preliminaries}
\subsection{Problem Definition}
In a programming scenario, let $\mathcal{U}$, $\mathcal{Q}$, and $\mathcal{C}$ denote the sets of $N$ learners, $M$ questions, and $K$ knowledge concepts, respectively. Each question typically tests one primary knowledge concept such as \textit{Array}. Each learner’s practice activities consist of a sequence of submitted codes and corresponding feedback. For learner $u$ at submission step $t$, he/she attempts to solve a question $q_{u,t} \in \mathcal{Q}$, drawn from a knowledge concept $c_{u,t} \in \mathcal{C}$, by writing and submitting code $x_{u,t}$, and receives feedback $r_{u,t}$, such as \textit{Wrong Answer} or \textit{Compile Error}. Feedback messages also reflect the learner's programming performance.
Existing PKT models typically represent $r_{u,t}$ as a binary value: $r_{u,t}=1$ indicates that learner $u$ received an correct result (i.e., solved the question $q_{u,t}$ with \textit{Accepted} feedback), while $r_{u,t}=0$ otherwise. Thus, each learner $u$ has a sequence of programming records:
\begin{equation}
l_{u} = { ({q}_{u,1}, {c}_{u,1}, {x}_{u,1}, {r}_{u,1}), \dots, ({q}_{u,T{u}}, {c}_{u,T{u}}, {x}_{u,T{u}}, {r}_{u,T{u}}) },
\end{equation}
where $T_{u}$ denotes the length of learner $u$'s programming sequence.

The evolution of the learner’s knowledge proficiency during the programming process is hard to record and quantify explicitly. Therefore, most previous approaches model PKT as a next-attempt performance prediction problem. The optimization objective is to minimize the cross-entropy loss between the predicted performance $\hat{r}_{u,t+1}$ (i.e., whether the learner will answer correctly on the next attempt) and the actual performance $r_{u,t+1}$ at attempt $t$~\cite{yin2023tracing}:
\begin{equation}
\label{eq:op}
\mathcal{L}_{u,t}^{pkt} = -r_{u,t+1} \log \hat{r}_{u,t+1} - \left(1 - r_{u,t+1}\right) \log \left(1 - \hat{r}_{u,t+1}\right).
\end{equation}
By jointly training on performance predictions using all programming data, the knowledge state of learners with respect to each question-related knowledge concept can be optimized.

In summary, we formally define the problem of Programming Knowledge Tracing (PKT) as follows:
\begin{definition}[\textbf{Programming Knowledge Tracing}]
Given the prior programming records of a learner $u$ up to and including submission step $t$, represented as a sequence $\{ ({q}_{u,1}, {c}_{u,1}, {x}_{u,1}, {r}_{u,1}), \ldots,$ $ ({q}_{u,t}, {c}_{u,t}, {x}_{u,t}, {r}_{u,t}) \}$, the goal is to: (1) estimate the internal programming knowledge state $h_{u,t}$ at attempt $t$ of this learner; (2) predict the performance $\hat{r}_{t+1}$ on the next question $q_{u,t+1}$.
\end{definition}

\begin{figure}
    \centering
    \scalebox{0.33}
    {\includegraphics{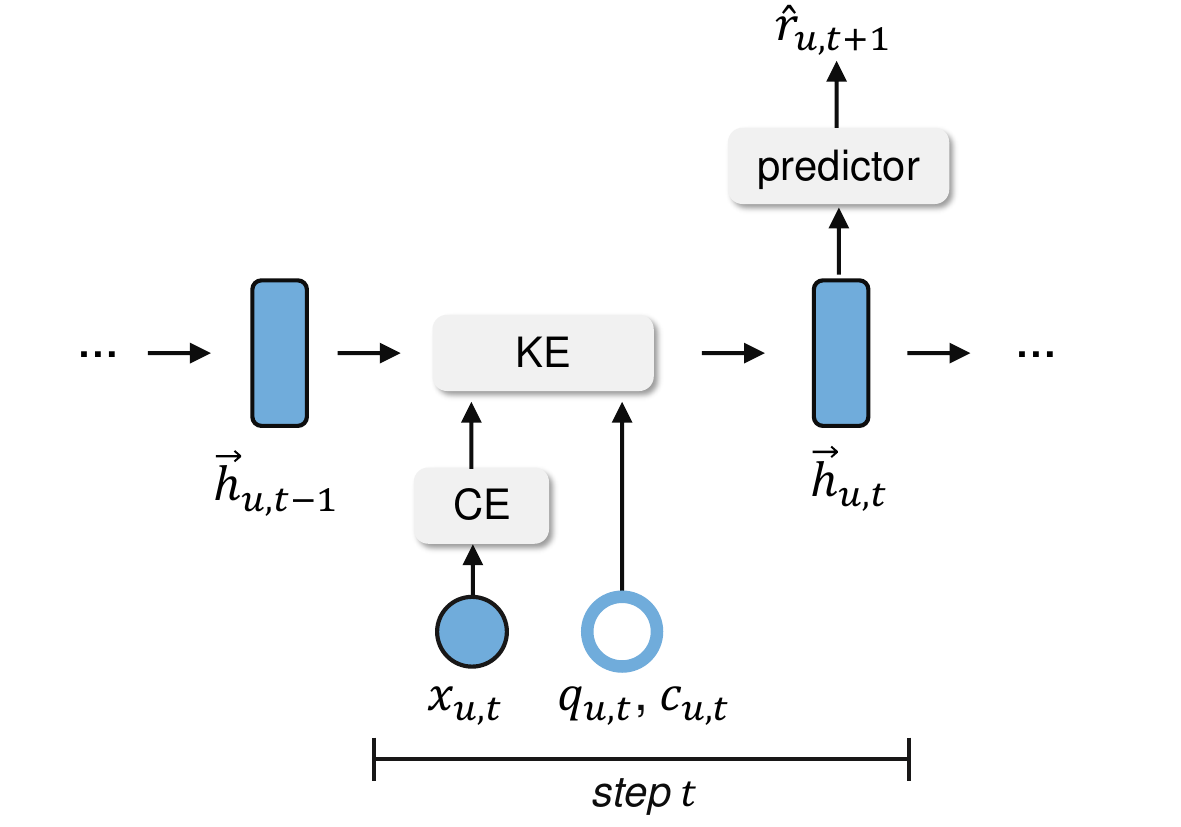}}
    \vspace{-0.3cm}
    \caption{The basic architecture of the PKT model.}
    \vspace{-0.5cm}
    \label{fig:pkt}
\end{figure}

\vspace{-0.2cm}
\subsection{PKT Backbone Model} \label{sec:pkt_atchitecture}
Our Coda framework needs to tune a trained PKT model (as the backbone model), such as DKT+~\cite{wang2017deep}, so this part introduces the general structure of the existing PKT model to serve as the basis for the \S~\ref{sec:coda} framework introduction.

Almost all of previous PKT models employ a sequential way, e.g., RNN, to recurrently model each coding submission of the learner.
Each recurrent unit typically consists of three basic components that work conjointly processing each coding submission once, until the end of the programming sequence: the \textbf{code encoder (CE)}, the \textbf{knowledge estimation (KE)} and the \textbf{predictor}, as illustrated in Figure~\ref{fig:pkt}.
At each attempt of the programming practice sequence (index by \textit{t}), \textbf{(1)}~\textbf{CE} first extracts and represents the code feature $\vec{x}_{u,t}$ from a learner $u$'s submission $x_{u,t}$ at attempt $t$, i.e., $\vec{x}_{u,t}={\rm CE}(x_{u,t})$.
Existing \textbf{CE} modules either extract semantic features regarding codes as a special language using NLP techniques~\cite{feng2020codebert,zhu2022programming} or first transform the code to a graph (e.g., abstract syntax tree~\cite{alon2018code2vec,li2022pst}) and then encode structural dependencies using graph modeling methods.
$\textbf{(2)}$~The \textbf{KE} component then estimates the knowledge state $\vec{h}_{u,t}$ for learner $u$ via the state $\vec{h}_{u,t-1}$ from the last attempt and the current submission record, i.e., $\vec{h}_{u,t+1}={\rm KE}(\vec{h}_{u,t-1},\vec{q}_{u,t},\vec{c}_{u,t},\vec{x}_{u,t},{r}_{u,t})$.
The embeddings $\vec{q}_{u,t}$ and $\vec{c}_{u,t}$ are usually obtained from their textual contents or learned through joint training~\cite{zhu2022programming}.
\textbf{(3)}~The \textbf{predictor} finally predicts the learner’s performance $\hat{r}_{u,t+1}$ to the next question ${q}_{u,t+1}$ based on the current knowledge state $\vec{h}_{u,t}$.


In our paper, we will not focus on the details of the pkt backbone model. The training and test details of a PKT backbone model are given in Appendix~\ref{app:algo}. Given a well-trained PKT backbone model, our Coda and bindings with it through its three key components described above.

\section{The Coda Framework}
\label{sec:coda}
\begin{figure*}
    \centering
    \scalebox{0.35}
	{\includegraphics{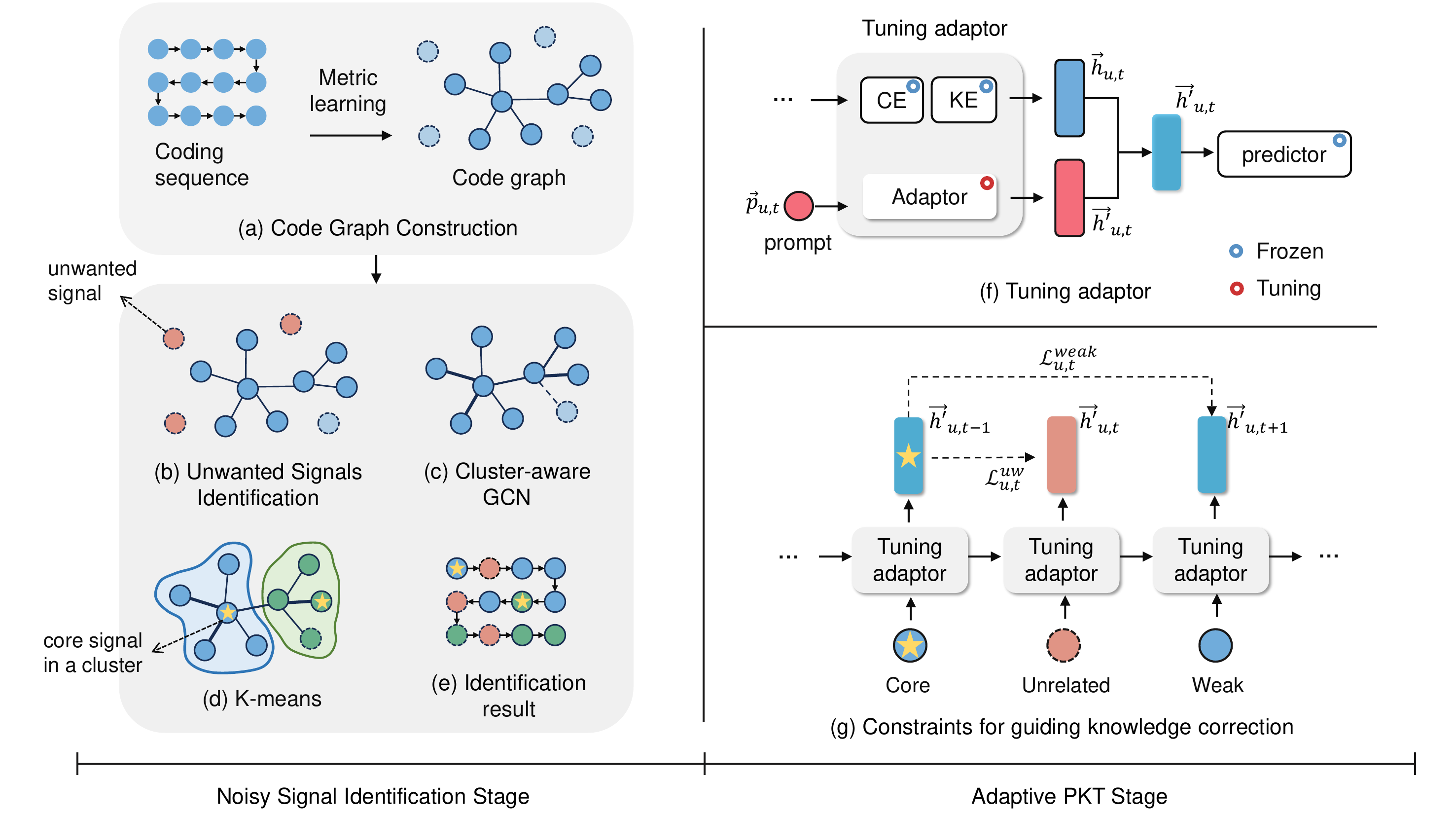}}
 \vspace{-0.3cm}
	\caption{The overall architecture of Coda. Left shows the stage for noisy signal identification, where (b) aims to find unwanted signals, and (c) and (d) perform weak signal fusion. Right shows the adaptive PKT stage with the tuning of Coda.}
 \vspace{-0.2cm}
	\label{fig:coda}
\end{figure*}

Coda is a general and model-agnostic framework designed to help existing PKT models recognize noisy programming submissions and correct compromised knowledge states. It can be integrated with PKT models in various ways, such as tuning or end-to-end joint training. For the sake of computational efficiency (as elaborated in \S~\ref{sec:intro}), we adopt the tuning approach. Specifically, we first train a conventional PKT model (e.g., DKT+~\cite{wang2017deep}) as the backbone. Then, we incorporate Coda into the backbone model and only train Coda's parameters while keeping the PKT backbone fixed. 
Figure~\ref{fig:coda} illustrates our proposed Coda framework comprising two stages: the \textit{noisy signal identification stage} (\S~\ref{sec:identify}) and the \textit{adaptive PKT stage} (\S~\ref{sec:adapt_pkt}). Assuming we have obtained a trained PKT backbone model, we will elaborate on each stage of Coda in detail.





\subsection{Noisy Signal Identification Stage} \label{sec:identify}
Identifying unwanted and weak noise signals accurately is extremely difficult from programming sequences because we lack a clear judging criterion or available labels regarding noisy submissions. Through the discussion of noisy submissions in \S~\ref{sec:intro}, we argue that noisy submissions and regular submissions from a specific learner exhibit distinct semantic characteristics: \textit{unwanted submissions differ significantly from normal submissions in terms of code contents, while weak submissions tend to resemble a certain core submission}. This stage conducts noise identification by exploiting semantic relations between codes submitted by a specific learner.

\subsubsection{\textbf{Code Graph Construction}}
To fully exploit the semantic relations between codes submitted by a specific learner, we convert the loose submitted code sequence of each learner into a tight code graph. The nodes on the graph represent the submitted code at each attempt, and the edges indicate the existence of semantic similarities between a pair of nodes. Intuitively, for a specific learner, unwanted code nodes may have very few edges even within a multi-hop range, and they might even be isolated. Meanwhile, a group of weak submission nodes may cluster around a central core node. Therefore, if we can construct an effective code graph for the learner's submitted code sequence, it becomes easier to identify and model noisy signals.

Inspired by previous works~\cite{chang2021sequential, li2021intention}, we propose an adaptive way based on metric learning to automatically construct code graph structures for each programming practice sequence to explore the distribution of different coding signals.
Technically speaking, this module aims to construct an undirected code graph $G_{u}=\{V_{u},A_{u}\}$ where $A_{u}\in\mathbb{R}^{T_{u}\times T_{u}}$ is the learnable adjacency matrix, and $V_{u}$ is the submitted code set of a learner. Each node $v_{u,i} \in V_{u}$ corresponds to the code $x_{u,i}$ at the attempting step $i$ within the programming sequence. The node embedding vector $\vec{x}_{u,i}\in\mathbb{R}^{d}$ is obtained from the CE (defined in \S~\ref{sec:pkt_atchitecture}) in a refined PKT backbone model.
Various graph construction techniques have been proposed to learn graph structure~\cite{xia2021graph,li2022mining,jin2023transferable}. Since the focus of this paper is not to devise more sophisticated techniques for graph learning, we use a simple yet effective similarity metric learning method~\cite{chang2021sequential,wang2020gcn,zhao2017physics} to transform the graph learning problem into node similarity evaluations.
Specifically, we first calculate a metric matrix $\mathbf{M}_{u}\in \mathbb{R}^{T_{u}\times T_{u}}$. Each of its element $M_{u,<i,j>}$ is calculated as:
\begin{equation}
    M_{u,<i,j>}={\rm sim}({\mathbf{W}}\odot\vec{x}_{u,i},{\mathbf{W}}\odot\vec{x}_{u,j}), \label{eq:M}
\end{equation}
where ${\rm sim}(\cdot,\cdot)$ is a metric function to calculate the similarity metric between code node $v_{u,i}$ and $v_{u,j}$ corresponding to the $i$-th and $j$-th submissions, respectively, in $u$'s programming sequence. It is flexible and can be implemented by various forms. $\mathbf{W}\in \mathbb{R}^{d}$ is a trainable vector to adaptively highlight different dimensions of the code embeddings $\vec{x}_{u,i}$ and $\vec{x}_{u,j}$. $\odot$ denotes the Hadamard product~\cite{kim2016hadamard}. We implement ${\rm sim}(\cdot,\cdot)$ using a weighted cosine similarity with the consideration of balancing expressiveness and complexity~\cite{chen2020iterative,liu2023guiding}.

After obtaining the metric matrix $\mathbf{M}_{u}$, it is necessary to normalize it to ensure non-negativity, considering that the cosine values $M_{u,<i,j>}$ fall within the range of $[-1, 1]$. Simply normalizing it without additional constraints may not regulate the graph sparsity, potentially resulting in a fully connected adjacency matrix~\cite{chang2021sequential,yue2024cooperative}. This may reduce the discrimination of relations between different code pairs, which is harmful to the observation of core and weak submissions, as well as limit the occurrence of isolated nodes, which is harmful to the observation of unwanted submissions. To better focus on the vital connections and disconnect the unwanted submissions, we adopt a ranking strategy~\cite{chang2021sequential} to control the sparsity of the graph. Specifically, we mask off those elements in $\mathbf{M}{u}$ that are smaller than a hyper-parameter by setting them to zero, which is obtained by ranking the metric value in $\mathbf{M}{u}$, to generate an informative adjacency matrix $A_{u}$, as follows:
\begin{equation}
    A_{u,<i,j>}=\begin{cases}
    1, \;\;\;\; M_{u,<i,j>} \ge \text{Rank}_{\epsilon}(\mathbf{M}_{u}) \\
    0, \;\;\;\; \text{otherwise}
    \label{eq:adj_m}
    \end{cases}
\end{equation}
where $\text{Rank}_{\epsilon}(\mathbf{M}_{u})$ returns the value of the $\epsilon$-th largest value in the metric matrix $\mathbf{M}_{u}$ to control the sparsity of the code graph.
Note that the structure of the code graph is learnable and changes continuously with the update of $\mathbf{W}$. The construction method is general, easy to implement, and able to perfectly cope with inductive learning (with new codes during test). Based on the code graph, we then conduct noise signal identification by exploiting similarity relations between nodes.

\subsubsection{\textbf{Unwanted Signals Identification}} \label{sec:unwanted}
This part aims to identify noisy signals originating from unrelated submissions. Intuitively, unrelated codes correlate weakly with other nodes within the code graph $G_{u}$ of a specific learner $u$. Therefore, they are most likely to be disconnected from other nodes within $G_{u}$ and become isolated nodes during the construction process.

However, it is difficult to determine unwanted submissions solely by relying on isolated nodes, as these may also represent novel or unique problem-solving approaches. To enhance the confidence in identifying unwanted nodes, we introduce an auxiliary view to exploit correlations between isolated code nodes and question-related solutions. The solutions refer to the correct submissions corresponding to the specific question. We argue that the correct solution must correspond to the practiced questions; otherwise, it will not pass the online judge. If a code node is similar to a certain solution, it is likely to be a useful and related submission signal for the current question.
Notably, codes with consistently low similarity to each solution related to the practiced question are not necessarily irrelevant to the question. These submissions may be relatively short and simple code snapshots that are question-related but require multiple modifications to solve the question correctly. At a glance, this idea cannot directly determine irrelevant signals. However, we can determine noisy signals by identifying those that are both isolated and question-uncorrelated simultaneously. Our basic assumption is that a submission that is both isolated and irrelevant to each solution is less relevant to the rest of the learner's programming sequence and the current question, as empirically verified by previous studies~\cite{liang2022help}.

Based on the above analysis, for each practiced question $q_{u,i}$ by $u$, we collect all the correct submissions (codes with the feedback \textit{Accepted}) from the entire training data corresponding to it as its solution set $\mathcal{S}_{q_{u,i}}$. We then calculate the similarity between the embeddings of isolated code $x_{u,i}$ and each solution code $x_{u,j'}$ as:
\begin{equation}
    \label{eq:code_solution}
    S_{i}=\{\{s_{i,j'}\}|j'\in\mathcal{S}_{q_{u,i}}\}, \;
    s_{i,j'}={\rm sim}({\mathbf{W}}\odot\vec{x}_{u,i},{\mathbf{W}}\odot\vec{x}_{u,j'}).
\end{equation}
The similarity evaluation function ${\rm sim}(\cdot,\cdot)$ is implemented using cosine similarity as Eq.~(\ref{eq:M}).
If the mean w.r.t. the distribution of $S_{i}$ is less than its median, we empirically consider the $x_{i}$ as an irrelevant submission.

After identifying the unwanted nodes, we remove them from the node set $V_{u}$. Each of the remaining isolated but question-related nodes needs to be added to the graph by linking the most similar node with it via cosine similarity score for further calculation. This process may result in multiple disjoint subgraphs (each containing at least two nodes) rather than necessarily forming a connected graph. In both cases, we uniformly denote them as $G_u$ for simplicity, as it will not affect subsequent calculations.

\subsubsection{\textbf{Cluster-aware Weak Signal Fusion}}
This section aims to distinguish strong signals from core code submissions and weak signals from minor modifications. In the context of programming, weak signals primarily stem from previous submissions with minor adjustments based on feedback information. These weak signals and their corresponding core submissions are more likely to be grouped into a cluster via their representations. Motivated by this idea, we first apply an alternative cluster-aware Graph Convolution Network (GCN) on the code graph to improve the discrimination among code nodes by gathering weak signals to strong ones.

The input of the cluster-aware GCN is a node embedding matrix $\{\vec{x}_{u,1},\vec{x}_{u,2},\cdots,\vec{x}_{u,n}\}$, where each $\vec{x}_{u,t}\in \mathbb{R}^{d}$ is the code $x_{u,t}$'s representation and $n<T_{u}$ is the number of nodes on the graph $G_{u}$ after removing unwanted nodes (i.e., the length of the learner programming sequence on a certain question besides unrelated submissions).
The output is a new node embedding matrix $\{\vec{x'}_{u,1},\vec{x'}_{u,2},\cdots,\vec{x'}_{u,n}\}$, where $\vec{x'}_{u,i}\in \mathbb{R}^{d'}$ with potentially different dimension $d'$. We iteratively conduct GCN $\textbf{Aggregate}(\cdot)$ with $L>0$ times as:
\begin{equation}
    \begin{split}
    \vec{x'}_{u,i}&= \vec{x}_{u,i}^{(L)}+\vec{x}_{u,i},\\
    \vec{x}_{u,i}^{(l)}&=\sigma\left(\mathbf{W_{a}}\cdot\mathbf{Aggregate}\left(\alpha_{i,j}\cdot\vec{x}_{u,j}^{(l-1)}; j\in\mathcal{N}_{i}\right)\right), l=1\sim L.
    \end{split}
    \label{eq:gcn}
\end{equation}
The above Eq.~(\ref{eq:gcn}) shows how a code node $v_{u,i}$ gathers the semantic signals from its neighboring set $\mathcal{N}_{u,i}$ as well its original embedding $\vec{x}_{u,i}$. $\vec{x}_{u,i}^{(0)}$ is defined as $\vec{x}_{u,i}^{(0)} = \mathbf{W}\odot\vec{x}_{u,i}$, where $\mathbf{W}$ is the same parameter as in Eq.~(\ref{eq:M}). $\mathbf{W_{a}}\in\mathbb{R}^{d\times d'}$ is the corresponding linear transformation’s weight matrix and $\sigma$ is the sigmoid activation function.
Note that $\mathbf{Aggregate}(\cdot)$ can be a function such as $\textit{Mean}$, $\textit{Sum}$, $\textit{Max}$, etc. We use the simple mean function here and leave other functions for future exploration.

To better identify whether the selected core node $v_{u,i}$ is the center of the cluster, we introduce a cluster-aware attention $\alpha_{i,j}$.
However, the cluster information of $v_{u,i}$ is not explicitly provided. To this end, we intuitively define the receptive field of cluster $c$ as $v_{u,i}$'s $k$-hop neighborhoods, with the assumption that each node $v_{u,i}$ in the graph is likely to be a core node forming a cluster $c$.
Then, we calculate a cluster-aware attention score using the embedding of each node in the assumed cluster and the cluster embedding $\vec{x}_{u,i}^{(c)}$ (i.e., the average value of all nodes’ embeddings in the cluster) as:
\begin{equation}
    \label{eq:attention}
    \alpha_{i,j} = \mathbf{Attention_{c}}\left(\mathbf{W_{c}}\odot\vec{x}_{u,i}\|\mathbf{W_{c}}\odot\vec{x}_{u,i}^{(c)}\|\mathbf{W_{c}}\odot\vec{x}_{u,j}; j\in\mathcal{N}_{u,i}^{(c)}\right),
\end{equation}
where $\mathbf{Attention_{c}}(\cdot)$ is implemented by a feed-forward neural network with a sigmoid activation function. $\|$ is the concatenation operator. $\mathcal{N}_{u,i}^{(c)}$ is the node set in the assumed cluster $c$, and $\mathbf{W_{c}}\in\mathbb{R}^{d}$ is a learnable transformation vector. Particularly, if $G_u$ contains several sub-graphs, the GCN operation should be separately conducted over each sub-graph to update node representations.

After obtaining cluster-aware code embeddings $\{\vec{x'}_{u,i}\}_{i=1}^{n}$, we divide them into several true clusters. 
Numerous clustering methods have already been proposed~\cite{xu2005survey}. In our implementation, a popular K-means algorithm~\cite{hartigan1979algorithm} is performed over $\{\vec{x'}_{u,1},\vec{x'}_{u,2},\cdots,\vec{x'}_{u,n}\}$, and generates $C_k$ clusters. The corresponding clustering index list is $\{idx_{u,1},idx_{u,2},\cdots,idx_{u,n}\}$.
In particular, the earliest submission in each cluster is considered the core signal and the remaining codes in the same cluster are identified as weak signals.

\subsection{Adaptive PKT Stage} \label{sec:adapt_pkt}
After performing the identification process, we can obtain the identifications of unwanted codes corresponding to unrelated submissions and the cluster indexes for each core signal from the significant submission and the weak signals from minor modifications.
Leveraging these noise indicators, this stage aims to improve the backbone PKT model under the noisy scenario as much as possible.

To facilitate the subsequent calculation, we transform each type of submission signal into the corresponding prompt $\vec{p}_u\in \mathbb{R}^{2d'}$.
Concretely, for the core submission with the code representation $\vec{x'}_{u,i}$ and each weak submission with the code representation $\vec{x'}_{u,j}$ in the same cluster $c$, let $\left[\mathbf{W_{p}}\odot\vec{x'}_{u,i}^{(c)}\|\vec{0}^{d'}\right]$ and $\left[\vec{0}^{d'}\|\mathbf{W_{p}}\odot\vec{x'}_{u,j}^{(c)}\right]$ be their prompts, respectively, where $\vec{0}^{d'}$ denotes a $d'$-dimensional zero vector.
For each unrelated submission with embedding $\vec{x'}_{u,r}$, its prompt is defined as $\left[\mathbf{W_{p}}\odot\vec{x'}_{u,r}^{(c)}\|\mathbf{W_{p}}\odot\vec{x'}_{u,r}^{(c)}\right]$. $\mathbf{W_{p}}\in\mathbb{R}^{d'}$ is a learnable transform vector.

\subsubsection{\textbf{Tuning Adaptor for PKT}}
Based on the prompt indicators, we employ a light and efficient tuning strategy to combine Coda with a refined original PKT model whose parameters are frozen, with the consideration of efficiency.


Inspired by LoRA~\cite{hu2021lora}, we implement a tuning adaptor using two trainable rank decomposition matrices, $\mathbf{W}_{A}\in\mathbb{R}^{b\times 2d'}$ and $\mathbf{W}_{B}\in\mathbb{R}^{b\times 2d'}$, which are integrated with the CE and KE modules in the backbone PKT model, as shown in Figure~\ref{fig:coda}~(f). The backbone PKT model's parameters are kept frozen. Specifically, the adaptor receives the identification prompt $\vec{p}_{u,t}$ at each step $t$ to sense different types of coding signals (including noisy signals and core submissions) and generates correction signals tailored to various contexts (i.e., the knowledge state and submission at step $t$). The correction signal is added to the original knowledge state $\vec{h}_{u,t}$ output from the refined KE component of the backbone PKT model to generate the corrected knowledge state as follows:
\begin{equation}
    \label{eq:fine_tune}
    \vec{h'}_{u,t}=\vec{h}_{u,t}+\vec{h'}_{u,t_{a}}, \;
    \vec{h'}_{t_{u,a}}=(\mathbf{W}_{A}^{T}\cdot\mathbf{W}_{B})\odot\vec{p}_{u,t},
\end{equation}
where $\vec{h'}_{u,t_{a}}$ is the correction signal. It adjusts the noisy state $\vec{h}_{u,t}$ at each dimension and generates a denoised knowledge state $\vec{h'}_{u,t}$.

To optimize the adaptor effectively, we summarize the characteristics of unwanted and weak signals as two mathematical constraints to control the knowledge state correction, as shown in Figure~\ref{fig:coda}~(g).
The first consideration is that unwanted signals are unrelated to programming practice and cannot reflect learners' actual knowledge states. To correct the biased knowledge states from these unrelated signals in the refined PKT model, we design a constraint that requires the knowledge state updated with irrelevant signal data to be similar to the knowledge state after the most recent question-related attempt (denoted as $t^{+}$). Related attempts include any core/weak submissions. This constraint is modeled with the following loss function:
\begin{equation}
    \label{eq:fine_tune_loss_1}
    \mathcal{L}_{u,t}^{uw}=D_{\text{KL}}\left(\vec{h'}_{u,t},\vec{h'}_{u,t^+}\right),
\end{equation}
where $D_{\text{KL}}(\cdot)$ denotes Kullback Leibler (KL) divergence~\cite{alemi2016deep} to close the gap between two knowledge states.
Note that $\mathcal{L}_{u,t}$ is omitted if there are no question-related attempts before the attempt step $t$.

For each weak submission $t$, the learner's knowledge state should not undergo significant changes and should be close to the state corresponding to a core submission. This can be formalized as:
\begin{equation}
    \label{eq:fine_tune_loss_2}
    \mathcal{L}_{u,t}^{weak}=D_{\text{KL}}\left(\vec{h'}_{u,t},\vec{h'}_{u,core}\right),
\end{equation}
where $\vec{h'}_{u,core}$ is the corrected knowledge state at core submission corresponding to the weak submission at the same cluster.

In summary, the tuning adaptor requires the following noise feature-based loss function:
\begin{equation}
    \mathcal{L}_{u,t}^{adptor}=\begin{cases}
    \mathcal{L}_{u,t}^{uw}, \;\;\;\; \text{submission}\;x_{u,t}\;\text{is unrelated}, \\
    \mathcal{L}_{u,t}^{weak}, \;\; \text{submission}\;x_{u,t}\; \text{is weak}.
    \label{eq:fine_tune_loss_coda}
    \end{cases}
\end{equation}

\subsubsection{\textbf{Navigational Regularization}}
To ensure that the adaptor accurately guides the backbone PKT model without adverse effects, we introduce an additional navigational regularization term to constrain the optimization. Specifically, using a Stochastic Gradient Descent (SGD) optimization strategy~\cite{ruder2016overview}, the model parameters are updated at each training batch. We propose that the model prediction accuracy after tuning at training batch $b$ should not be worse than the last batch $(b-1)$. Let $\theta_{b}$ denote the model parameters after the $b$-th batch; this regularization term can be formalized as:
\begin{equation}
    \label{eq:navigation_loss}
    \begin{split}
        \mathcal{L}^{nav}_{b}&=\left|\sum_{u=1}^{N}\sum_{t=1}^{T_{u}}\mathcal{L}_{u,t}^{pkt}(\theta_{b})-\sum_{u=1}^{N}\sum_{t=1}^{T_{u}}\mathcal{L}_{u,t}^{pkt}(\theta_{b-1})\right|,
    \end{split}
\end{equation}
where $\mathcal{L}_{u,t}^{pkt}(\theta_{b})$ denote the prediction loss on one training data $({q}_{u,t},{c}_{u,t},{x}_{u,t},{r}_{u,t})$ based on the updated model parameter $\theta_{b}$. $|\cdot|$ is the absolute value operator.
The Eq.~(\ref{eq:navigation_loss}) is intuitive, but computationally complex, and it requires a prediction based on all the training data at each batch.
Therefore, we give the upper bound of Eq.~(\ref{eq:navigation_loss}) with the consideration of the computational complexity and storage space in actual implementation based on \textbf{Property 1}:

\textbf{Property 1.}
$\mathcal{L}^{nav}_b$ \textit{is bounded as follows:}
\begin{equation}
    \begin{split}
    \mathcal{L}^{nav}_b \leq \left|\left(\nabla_{\theta_{b-1}}\mathcal{L}^{pkt}(\theta_{b-1})\right)^{T}\right| \cdot\left|\delta \theta_{b-1}\right|+\left|\delta \theta_{b-1}\right|^T \cdot|J^{T}J| \cdot\left|\delta \theta_{b-1}\right|,
    \label{eq:l_nav_bound}
    \end{split}
\end{equation}
where $\nabla_{\theta_{b-1}}\mathcal{L}^{pkt}(\theta_{b-1})$ is the training gradient after the $(b-1)$-th batch tuning, $\delta\theta_{b-1}$ denote the parameter change between the $b$-th batch and $(b-1)$-th batch,i.e., $\delta\theta_{b-1}=\theta_{b}-\theta_{b-1}$. $J$ is the Jacobian matrix~\cite{wilamowski2010improved}. In implement, the Jacobian matrix $J$ is denoted using the training gradient $\nabla_{\theta_{b-1}}\mathcal{L}^{pkt}(\theta_{b-1})$.
Compared to Eq.~(\ref{eq:navigation_loss}), the Eq.~(\ref{eq:l_nav_bound}) is calculated once per training batch, requiring additional storage space for storing model parameters and gradients, which is acceptable in practice.
See the Appendix~\ref{app:proof} for the full proof.

\subsection{\textbf{Optimization}}
Overall, at each training batch $b$ with programming records of a batch of learners $\mathcal{U}_b \sim \mathcal{U}$, putting the prediction objective (Eq.~(\ref{eq:op})), the tuning objective (Eq.~(\ref{eq:fine_tune_loss_coda})) and the navigation regularization (Eq.~(\ref{eq:l_nav_bound})) together, we have the following training loss function:
\begin{equation}
    \mathcal{L}^{Coda}=\sum_{\mathcal{U}_b \sim \mathcal{U}}\left(\sum_{u \in \mathcal{U}_b}\sum_{t=1}^{T_{u}}\left(\mathcal{L}_{u,t}^{pkt}+\mathcal{L}_{u,t}^{adaptor}\right)+\mathcal{L}^{nav}_b\right).
    \label{eq:coda_loss}
\end{equation}

\subsection{Computational Complexity}
During the training phase, the code graph is constructed only once for each learner $u$'s sequence $l_u$. The primary computational complexity arises from calculating the adjacency matrix (see Eq.~(\ref{eq:M}) and Eq.~(\ref{eq:adj_m})), which has a complexity of $O(T_u \times T_u)$, where $T_u$ is the length of the programming sequence. When identifying irrelevant noise, we introduce an auxiliary task with a complexity of $O(w)$, where $w$ represents the average number of correct submissions per question. Identifying weak submissions mainly involves the GCN, with a complexity of approximately $O(T_u \times L)$, where $L$ is the number of layers in the GCN, and the clustering complexity is $O(T_u \times C_k)$, where $C_k$ is the number of clusters. When tuning the backbone PKT model, the primary complexity arises from predicting the learner's performance for each submission step, with a complexity of $O(T_u)$. Therefore, the overall complexity of modeling each learner during the entire training phase is $O(T_u \times T_u) + O(w) + O(T_u \times L) + O(T_u \times C_k) + O(T_u)$.
During the test phase, for a learner $u$, the code graph needs to be constructed at each step, resulting in a complexity of $O(\sum_{i=1}^{T_u} i^2) + O(T_u \times w) + O(\sum_{i=1}^{T_u} i \times L) + O(\sum_{i=1}^{T_u} i \times C_k) + O(\sum_{i=1}^{T_u} i)$.


\section{Experiments}
We conduct comprehensive experiments to address the following research questions (RQs): \textbf{RQ1} Can Coda enhance the accuracy of knowledge state estimation for backbone PKT models, thereby improving the prediction of student performance? \textbf{RQ2} How do the various components designed in Coda influence its performance? \textbf{RQ3} What is the effectiveness of code graph modeling? \textbf{RQ4} How powerful is Coda in tracing the evolution of knowledge states?

\subsection{Datasets}

\begin{table*}[t]
\caption{Performance comparison. The best performance is highlighted in bold, and $\uparrow$ ($\downarrow$) means the higher (lower) score the better (worse) performance, the same as below.}
\vspace{-0.3cm}
\setlength{\tabcolsep}{0.7mm}{
\scalebox{0.79}{

\begin{tabular}{c|c|ccc|cccccc|ccccc}
\hline
\multirow{2}{*}{Datasets} & \multirow{2}{*}{Metrics} & \multicolumn{3}{c|}{KT Models} & \multicolumn{6}{c|}{PKT Models} & \multicolumn{5}{c}{Coda} \\ \cline{3-16}

 & & DKT & AKT & LPKT & DKT+      & PDKT        & Code-DKT        & Help-DKT              & PST   & ECKT    & Coda-DKT+   & Coda-PDKT    & Coda-Code-DKT   & Coda-Help-DKT & Coda-PST \\ \hline

\multirow{3}{*}{BePKT} & AUC  (\%) $\uparrow$

& 52.77 & 54.12 & 53.69 &  53.96  &  55.32                    &56.14                            &60.10                &- &59.63      &{56.34}    &{59.74}      &{59.42}      &\textbf{62.31}  &-         \\

& F1-score  (\%) $\uparrow$   &9.30&12.51&10.07   &10.13                      &14.68                        &21.03                       &31.65                              &-    &34.32     &{26.96}   &{31.39}      &{34.22}   &\textbf{36.03}   &-          \\

& RMSE  (\%) $\downarrow$  &44.67&30.59&31.78 &42.93                         &40.01                   &34.56                   &$31.96    $                       &-    &35.44   &{41.24}       &${39.03}$        &${34.25}$   &$\textbf{31.33} $     &-         \\ \hline

\multirow{3}{*}{AtCoder} & AUC  (\%) $\uparrow$

&52.37&53.34&54.62 & 52.46 &  53.57                  &57.23                          &58.32              &-   &59.32      &{54.02}        &{59.03}  &$\textbf{62.96}$        &{59.63}   &-         \\

& F1-score  (\%) $\uparrow$   &7.00 &15.96&22.74   &7.02                      &16.39                 &27.42                         &31.44                             &-       &32.40      &{16.31}     &{34.69}   &$\textbf{35.37} $      &33.21   &-          \\
& RMSE  (\%) $\downarrow$   &45.17&41.88&38.82 &44.37                        &39.66                  &31.63                   &$39.42$                       &-   &37.54       &43.20   &38.27       &$\textbf{30.32}$  &38.36     &-        \\ \hline

\multirow{3}{*}{BePKT\_C++} & AUC  (\%) $\uparrow$

&52.96&54.17&53.29 & 54.17  &  55.29                    &55.63                             &60.58                &60.21   &60.02    &{57.89}        &{59.30} &{58.83}    &{62.97}  &\textbf{63.75}          \\
& F1-score  (\%) $\uparrow$ &8.37&10.24&10.01 &10.32                  &14.29                 &18.21                       &33.75                          &28.57   &27.92            &{27.16}     &{30.65}      &33.49           &{37.19}  &$\textbf{37.40}$          \\
& RMSE  (\%) $\downarrow$ &44.28&43.77&43.82 &42.09                         &40.66                      &34.08           &31.30                        &31.07   &32.03     &40.89     &39.57       &33.65 &30.97    &$\textbf{30.20}$          \\
\hline

\multirow{3}{*}{AtCoder\_C} & AUC  (\%) $\uparrow$ &52.51&53.07&53.27 &$53.12$                       &$54.14$                  &$56.54$                          &$59.83$                          &$64.17$  &$64.43$       &${54.64}$        &${57.36}$    &${62.14}$   &${60.62}$       &$\textbf{65.94}$          \\
& F1 (\%) $\uparrow$ &7.84&10.67&12.42 &$8.49$                     &$12.33$                      &$24.99$                   &$33.89$                   &  $51.74$ &$52.04$    &${18.31}$  &${32.33}$          &${40.67}$     &${36.08}$     &$\textbf{61.28}$          \\

& RMSE   (\%) $\downarrow$ &43.20&42.96&41.77 & 42.20                    &  $41.30$              &  $38.41$               &  $38.03     $                     &  $40.80$ & $40.01$    &${41.11}$      &${40.93}$           &${38.28}$   &$\textbf{37.90}$     &${39.26}$ \\ \hline
\end{tabular}
}}
\label{tab:main}
\end{table*}
We evaluate the effectiveness of Coda using four real-world public datasets: BePKT\footnote{\url{https://drive.google.com/drive/folders/1Jt6f0MV1paGLlctJqxHtdF1Vh2mUnsoV?usp=sharing}}, AtCoder\footnote{\url{https://github.com/IBM/Project_CodeNet}}, BePKT\_C++ and AtCoder\_C.
These datasets contain learner programming records, including submitted code and feedback on judgments, as well as question-knowledge correlations where each question is linked to a specific knowledge concept. Specifically, BePKT and AtCoder include multiple programming languages (e.g., C and C++), while BePKT\_C++ and AtCoder\_C are subsets of BePKT and AtCoder, respectively, that focus exclusively on C++ and C submissions. The reason why we construct the two subsets is the baseline (i.e., PST~\cite{li2022pst} introduced in \S~\ref{sec:baseline}) only supports the single language programming scenario.
The complete statistics of these datasets are summarized in Table~\ref{tab:app_dataset_static}.
Detailed dataset information and preprocessing steps are provided in Appendix~\ref{app:dataset}.

\begin{table}[t]
\centering
\vspace{-0.3cm}
\caption{The statistics of four programming datasets.}
\vspace{-0.3cm}
\setlength{\tabcolsep}{0.7mm}{
\scalebox{0.8}{
\begin{tabular}{l|rrrr}
\hline
{Dataset} & {BePKT} & {AtCoder} & {BePKT\_C++} & {AtCoder\_C} \\ 
\midrule
\#learners & 479 & 5,523 &447 & 5,416 \\
\#questions  & 423 & 832 &146 & 832 \\
\#knowledge concepts & 83 & 832 & 83 & 832 \\
\#submissions & 27,679 & 197,419 & 4,278 & 27,297     \\
\#submissions per learner & 57.78 & 35.74 & 9.57 & 5.04 \\ \hline
\end{tabular}
}}
\vspace{-0.5cm}
\label{tab:app_dataset_static}
\end{table}
\vspace{-0.2cm}
\subsection{Experimental Setups}
\subsubsection{\textbf{Baselines}}
\label{sec:baseline}
For baselines, we first select some traditional knowledge tracing models that are classic but not specifically tailored for programming scenarios, i.e., DKT~\cite{piech2015deep}, AKT~\cite{ghosh2020context}, and LPKT~\cite{ShenS021lpckt}.
Then, we select several well-adopted PKT models, i.e., DKT+~\cite{wang2017deep}, PDKT~\cite{zhu2022programming}, Code-DKT~\cite{shi2022code}, Help-DKT~\cite{liang2022help}, PST~\cite{li2022pst} and ECKT~\cite{yu2024eckt}. Notably, PST requires running in a single programming language scenario (i.e., the BePKT\_C++ and AtCoder\_C datasets). 
The descriptions of baselines are listed as follows:
\begin{itemize}[leftmargin=10pt]
    \item \textbf{DKT} leverages RNN to assess knowledge mastery by iteratively predicting learner next performance.
    \item \textbf{AKT} is the context-aware attentive KT model. It uses the two self-attentive encoders to learn context-aware representations of the exercises and answers.
    \item \textbf{LPKT} monitors learners' knowledge state through directly modeling their learning process with a basic chain: \textit{exercise—answer time—answer}.
    \item \textbf{DKT+} uses abstract syntax trees (AST)~\cite{alon2018code2vec,li2022pst} to represent programs and uses recurrent neural networks to track learners' knowledge levels.
    \item \textbf{PDKT} uses a double-sequence model with exponential decay attention to model problem and code sequences.
    \item \textbf{Code-DKT} constructs the AST and adopts both node embedding and path embedding to effectively model the problem text.
    \item \textbf{Help-DKT} introduces an interpretable cognitive model that integrates programming error classification as a conceptual indicator into a personalized feedback-based matrix, aiming to estimate student ability at the conceptual level.
    \item \textbf{PST} builds a code information graph and a code tracing graph containing rich structural and semantic information from code contents. Additionally, it takes into account code modification content between consecutive submissions to capture knowledge state changes. 
    \item \textbf{ECKT} integrates code features with questions' knowledge concepts and descriptions generated by an LLM from codes. In this paper, we generate knowledge concept embeddings based on the original datasets and generate each question description embedding by mean-pooling all codes related to the question.
\end{itemize}


\subsubsection{\textbf{Evaluation Metrics}}
As knowledge states cannot be directly observed in practice, it is common to indirectly assess PKT tasks through predicting learner response performance on test datasets~\cite{li2022pst,gao2024zero,li2025foundation}.
To evaluate model performance, we adopt ACC, AUC and RMSE as metrics from the perspectives of classification and regression following the~\cite{gao2021rcd,yao2024adard}.

\subsubsection{\textbf{Implementation Details}}
We implement our Coda framework with five well-adopted PKT backbone models, i.e., DKT+~\cite{wang2017deep}, PDKT~\cite{zhu2022programming}, Code-DKT~\cite{shi2022code}, Help-DKT~\cite{liang2022help} and PST~\cite{li2022pst}.
We call them Coda-X, e.g., Coda-PDKT and Coda-Help-DKT.
We split 70\%, 10\% and 20\% programming sequences of each dataset as the training set, the valid set, and the test set, respectively.
We utilize a pre-trained Code-Bert\footnote{\url{https://github.com/microsoft/CodeBERT}} to encode input codes for each baseline with the setting of dimension $d$ is 768.
The setups for training PKT backbone models refer to the original papers and each backbone is trained on training data. 
We implement our Coda-X models by tuning these refined baselines.
The parameter $\epsilon$ in Eq.~(\ref{eq:adj_m}) is sampled from the ``0.2/0.4/0.5/0.6/0.8 of the square of the sequence length" to regulate the sparsity of the code graph.
The GCN layers $L$ for Eq.~(\ref{eq:gcn})
and cluster-aware hop $k$ in Eq.~(\ref{eq:attention}) are both 2, following the related works~\cite{chang2021sequential}.
The number of clusters of k-means $C_k$ is searched in [1, 3, 5].
The dimension $d'$ is equal to the dimension $d$ of original code embedding and $b$ in Eq.~(\ref{eq:fine_tune}) is $d'/2$.
Furthermore, we set the mini-batch size as 32 and the learning rate is searched in $[1e^{-3}, 1e^{-2}, 1e^{-1}]$ for each model.
For training, all network parameters are initialized with Xavier initialization~\cite{glorot2010understanding}.
Each model is implemented by PyTorch~\cite{paszke2019pytorch} and optimized by Adam optimizer ~\cite{kingma2014adam}.
All experiments are run on a cluster of Linux servers with NVIDIA TITAN V100 GPUs.
Code is available at \url{https://github.com/bigdata-ustc/Coda-PKT}.

\subsection{Next Attempt Performance Prediction (RQ1)}
To evaluate model effectiveness, we compare all methods based on learners' subsequent attempt prediction accuracy, as depicted in Table~\ref{tab:main}. We can see that:
(1) Compared to traditional KT models, the PKT models and Coda-Xs show superior performance across all metrics. Traditional KT models are designed for a single submission per problem and ignore code information, leading to biases in scenarios with multiple submissions.
(2) In comparison to PKT models, our Coda-Xs demonstrate better prediction accuracy. For example, on the BePKT\_C++ dataset, Coda improves Help-DKT and PST by 3.44\% and 8.83\% in terms of F1 score accuracy, respectively. It reveals that our Coda to denoising programming sequences is both justified and effective.
(3) Most models perform better on two single language datasets, suggesting that longer programming sequences have more noise, thus limiting model performance.

\subsection{Investigation of Key Components (RQ2)}
\begin{table}
    \setlength\tabcolsep{1.pt}
    \centering
    \caption{Ablation study on BePKT\_C++.}
    \vspace{-0.3cm}
    \renewcommand\arraystretch{0.8}
    \setlength{\tabcolsep}{0.8mm}{
    \scalebox{0.8}{
        \begin{tabular}{lccc}
        \hline
        {Methods} & \;\;{AUC (\%) $\uparrow$} & {F1 (\%) $\uparrow$} & {RMSE (\%) $\downarrow$} \\
        \midrule
        Coda-PST & \textbf{63.75} & \textbf{37.40} & \textbf{30.20} \\
        \midrule
        w/o iso & {62.57} & {36.92} & {30.60} \\
        w/o sim & {63.32} & {37.23} & 30.63 \\
        w/o uw & 62.21 & 37.02 & 31.43 \\
        w/o gcn & 63.00 & 36.92 & 30.74 \\
        w/o weak & 61.58 & 34.75 & 32.30 \\
        w/o nav & 63.60 & 37.34 & 30.34 \\
        \hline
        \end{tabular}
    }}
    \label{tab:variants}
    \vspace{-0.0cm}
\end{table}

\begin{figure*}
    \centering
    \scalebox{0.85}
	{\includegraphics{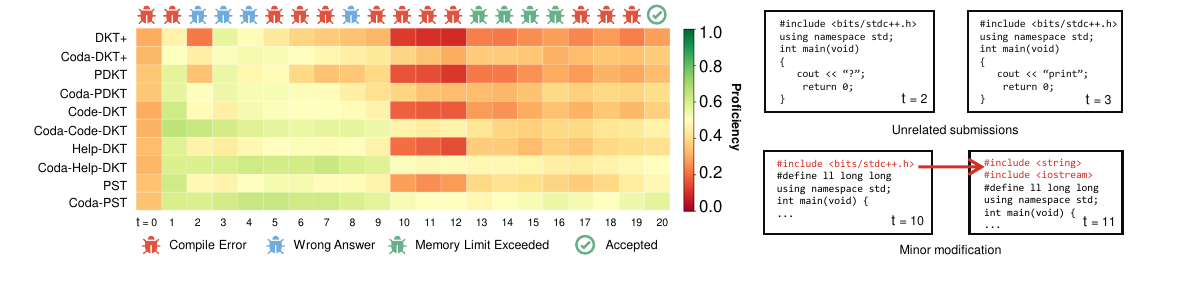}}
    \vspace{-0.75cm}
	\caption{Visualization of Code-X and original PKT models tracing knowledge of a learner along the practice sequence.}
	\label{fig:state}
    \vspace{-0.3cm}
\end{figure*}
To validate the effectiveness of each Coda design, we conduct an ablation study on Coda-PST using the BePKT\_C++ dataset. Specifically, we implement various Coda variants including ``w/o~iso'' by ignoring the isolated nodes (\S~\ref{sec:unwanted}), ``w/o~sim'' by removing the similarity calculation between codes and solutions (i.e., Eq.~(\ref{eq:code_solution})), ``w/o~uw'' by discarding the identification of unwanted signals (i.e., Eq.~(\ref{eq:fine_tune_loss_1})), ``w/o~gcn'' by skipping the clustering-aware GCN (i.e., Eq.~(\ref{eq:gcn})), ``w/o~weak'' by discarding the identification of weak signals (i.e., Eq.~(\ref{eq:fine_tune_loss_2})), and ``w/o~nav'' by discarding the navigational regularization (i.e., Eq.~(\ref{eq:l_nav_bound})).
Table~\ref{tab:variants} lists the prediction performance. From the table, we can see that each component of Coda significantly enhances the predictive performance of the PKT task.



\subsection{Analysis of the Code Graph (RQ3)}

\begin{figure}
    \centering
    \scalebox{0.65}
	{\includegraphics{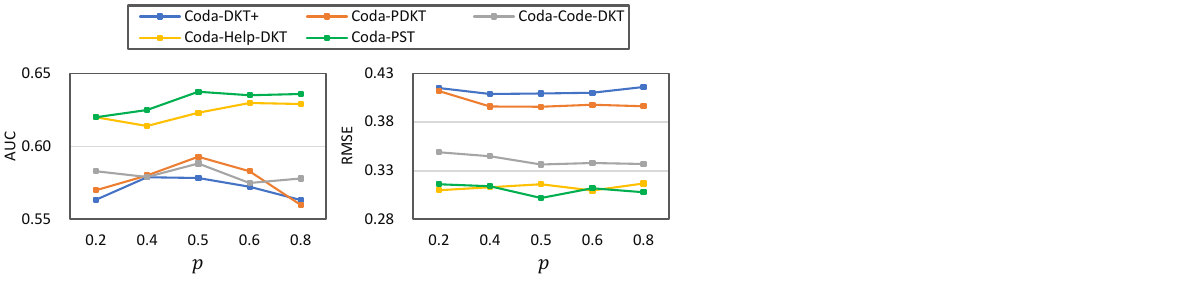}}
 \vspace{-0.9cm}
	\caption{Performance under different sparse levels of the code graph on BePKT\_C++.}
	\label{fig:size}
 \vspace{-0.45cm}
\end{figure}

The parameter $\epsilon$ in Eq.(\ref{eq:adj_m}) determines both the sparsity of the constructed code graph and the subsequent prediction accuracy.
To explore the impact of varying graph sparsity, we train the Coda-Help-DKT on several sparse code graphs where $\epsilon$ is set to ``$p$ of the square of the sequence length", with $p \in \{0.2, 0.4, 0.5, 0.6, 0.8\}$.
Figure~\ref{fig:size} illustrates the prediction scores for different sparse setups on the BePKT\_C++ dataset, clearly demonstrating that the model's performance improves with increasing $\epsilon$ when $\epsilon$ is small. However, once $\epsilon$ exceeds a certain threshold (e.g., $0.5$ of the square of the sequence length for Coda-PDKT), the performance gains become less pronounced.

\subsection{Visualization of Knowledge Tracing (RQ4)}
One of the most practical aspects of PKT is its ability to diagnose proficiency of specific knowledge concepts for learners at each attempt during programming.
To gain deeper insights into how Coda tracks the evolution of programming knowledge states, this section illustrates the tracing process using various models across a learning sequence.
Specifically, for a trained PKT model, we extract the knowledge proficiency on question-related concept $c$ from the $c$-th dimensional element in the knowledge state $\vec{h}_{u,t}$ inspired by~\cite{liu2019ekt,yin2023tracing}. We randomly sample a real programming sub-sequence of a learner ($id=320$) response to the question ($id=1304$) that tests ``\textit{Addition of Large Numbers}'' from BePKT\_C++, and visualize the evolution of proficiency on the concept tested by the question. Figure~\ref{fig:state} shows the tracing process learned by several models.

From the visualization, it is obvious that all Coda-X models demonstrate more stable knowledge tracing after the initial state ($t=0$) compared to original PKT models. The baselines may be susceptible to noise stemming from noisy submissions. Upon observing the learner's submitted codes (presented in the right part of Figure~\ref{fig:state}), it is evident that baselines are misled by noisy signals. For instance, at $t=2$ and $t=3$, the learner's codes are unrelated to the question. Coda-Xs effectively identifies noise submissions, thus preventing an underestimation of student knowledge due to these buggy examples. In contrast, PKT baselines are misled by irrelevant data. Similarly, at $t=10$ and $t=11$, the learner makes minor code modifications, indicating that his/her knowledge state should be relatively stable. While these submissions have no explicit bugs, they are redundant to PKT models. Coda-Xs remain unaffected by these minor changes due to its ability to identify and correct noisy signals, while the original PKT models continue to be misled, consistently underestimating learner knowledge proficiency.

\section{Related Work}
\label{sec:related_work}
In the field of educational data mining, Programming Knowledge Tracking (PKT) has taken on a significant role, focusing on the assessment.
PKT aims to dynamically diagnose learners' proficiency on specific programming knowledge. It extends traditional knowledge tracking (KT) techniques~\cite{piech2015deep,ghosh2020context,ShenS021lpckt,zhang2024understanding,wang2024survey,ma2024hd} and provides important technology for diagnosing students' coding abilities by integrating code features related to open programming environments. Existing PKT methods mainly consist of three key modules: code encoder (CE), knowledge estimation (KE), and predictor. For example, DKT+~\cite{wang2017deep} first uses Code2Vec technology to encode the code submitted by students, then models the student's answer sequence using LSTM~\cite{hochreiter1997long}, and uses the final representation to model the student's programming ability status. PDKT~\cite{zhu2022programming} incorporates representations of questions and knowledge points on this basis to enhance the capabilities of the knowledge modeler. Code-DKT~\cite{shi2022code} utilizes attention mechanisms to automatically extract and select domain-specific code features to extend DKT. HELP-DKT~\cite{liang2022help} combines students' original code encoding with error category information to enhance the capabilities of the knowledge modeler. PST~\cite{li2022pst} models differences between students' submitted codes in order to model their programming skills and knowledge, achieving state-of-the-art performance. ECKT~\cite{yu2024eckt} integrates code features with questions' knowledge concepts and descriptions generated by a large language model (LLM) from the code. GPKT~\cite{wu2024programming} learns informative code and question representations from a code-question-knowledge network for better prediction.  However, all these methods are limited by noise in students' answer sequences, which affects their applicability in real-world environments.

\section{Conclusion and Future Research}
In this study, we aim to address the prevalent challenges in programming knowledge tracing (PKT), with a particular focus on the impact of noisy signals encountered during long-term programming activities. We identify two prominent types of noisy signals: unwanted signals from unrelated submissions and weak signals from minor modifications. Both types significantly hinder the accurate representation of a learner's knowledge state.
To mitigate the impact of noise on PKT models, we propose the Coda framework. This framework enhances existing PKT solutions by enabling them to identify noisy signals and adaptively reduce their negative effects. Our approach involves transforming code sequence modeling issues into code graph modeling problems by leveraging the semantic relationships in the code. Additionally, we introduce a lightweight tuning adapter to seamlessly integrate with the PKT task.
Extensive experiments on four real-world datasets validate the effectiveness of our Coda model. Future research will identify additional types of noise and incorporate specific buggy feedback to achieve more comprehensive denoising for the programming knowledge tracing task.

\begin{acks}
This research was partially supported by grants from the National Natural Science Foundation of China (Grants No. 62337001), the Key Technologies R \& D Program of Anhui Province (No. 202423k09020039), and the Fundamental Research Funds for the Central Universities.
\end{acks}

\newpage

\bibliographystyle{ACM-Reference-Format}
\bibliography{sample-base}


\begin{thebibliography}{55}


\ifx \showCODEN    \undefined \def \showCODEN     #1{\unskip}     \fi
\ifx \showISBNx    \undefined \def \showISBNx     #1{\unskip}     \fi
\ifx \showISBNxiii \undefined \def \showISBNxiii  #1{\unskip}     \fi
\ifx \showISSN     \undefined \def \showISSN      #1{\unskip}     \fi
\ifx \showLCCN     \undefined \def \showLCCN      #1{\unskip}     \fi
\ifx \shownote     \undefined \def \shownote      #1{#1}          \fi
\ifx \showarticletitle \undefined \def \showarticletitle #1{#1}   \fi
\ifx \showURL      \undefined \def \showURL       {\relax}        \fi
\providecommand\bibfield[2]{#2}
\providecommand\bibinfo[2]{#2}
\providecommand\natexlab[1]{#1}
\providecommand\showeprint[2][]{arXiv:#2}

\bibitem[Alemi et~al\mbox{.}(2016)]%
        {alemi2016deep}
\bibfield{author}{\bibinfo{person}{Alexander~A Alemi}, \bibinfo{person}{Ian Fischer}, \bibinfo{person}{Joshua~V Dillon}, {and} \bibinfo{person}{Kevin Murphy}.} \bibinfo{year}{2016}\natexlab{}.
\newblock \showarticletitle{Deep variational information bottleneck}.
\newblock \bibinfo{journal}{\emph{arXiv preprint arXiv:1612.00410}} (\bibinfo{year}{2016}).
\newblock


\bibitem[Alon et~al\mbox{.}(2018)]%
        {alon2018code2vec}
\bibfield{author}{\bibinfo{person}{Uri Alon}, \bibinfo{person}{Meital Zilberstein}, \bibinfo{person}{Omer Levy}, {and} \bibinfo{person}{Eran Yahav}.} \bibinfo{year}{2018}\natexlab{}.
\newblock \bibinfo{title}{code2vec: Learning Distributed Representations of Code}.
\newblock
\showeprint[arxiv]{1803.09473}~[cs.LG]


\bibitem[Chang et~al\mbox{.}(2021)]%
        {chang2021sequential}
\bibfield{author}{\bibinfo{person}{Jianxin Chang}, \bibinfo{person}{Chen Gao}, \bibinfo{person}{Yu Zheng}, \bibinfo{person}{Yiqun Hui}, \bibinfo{person}{Yanan Niu}, \bibinfo{person}{Yang Song}, \bibinfo{person}{Depeng Jin}, {and} \bibinfo{person}{Yong Li}.} \bibinfo{year}{2021}\natexlab{}.
\newblock \showarticletitle{Sequential recommendation with graph neural networks}. In \bibinfo{booktitle}{\emph{Proceedings of the 44th international ACM SIGIR conference on research and development in information retrieval}}. \bibinfo{pages}{378--387}.
\newblock


\bibitem[Chen et~al\mbox{.}(2020)]%
        {chen2020iterative}
\bibfield{author}{\bibinfo{person}{Yu Chen}, \bibinfo{person}{Lingfei Wu}, {and} \bibinfo{person}{Mohammed Zaki}.} \bibinfo{year}{2020}\natexlab{}.
\newblock \showarticletitle{Iterative deep graph learning for graph neural networks: Better and robust node embeddings}.
\newblock \bibinfo{journal}{\emph{Advances in neural information processing systems}}  \bibinfo{volume}{33} (\bibinfo{year}{2020}), \bibinfo{pages}{19314--19326}.
\newblock


\bibitem[Feng et~al\mbox{.}(2020)]%
        {feng2020codebert}
\bibfield{author}{\bibinfo{person}{Zhangyin Feng}, \bibinfo{person}{Daya Guo}, \bibinfo{person}{Duyu Tang}, \bibinfo{person}{Nan Duan}, \bibinfo{person}{Xiaocheng Feng}, \bibinfo{person}{Ming Gong}, \bibinfo{person}{Linjun Shou}, \bibinfo{person}{Bing Qin}, \bibinfo{person}{Ting Liu}, \bibinfo{person}{Daxin Jiang}, {and} \bibinfo{person}{Ming Zhou}.} \bibinfo{year}{2020}\natexlab{}.
\newblock \bibinfo{title}{CodeBERT: A Pre-Trained Model for Programming and Natural Languages}.
\newblock
\showeprint[arxiv]{2002.08155}~[cs.CL]


\bibitem[Gao et~al\mbox{.}(2021)]%
        {gao2021rcd}
\bibfield{author}{\bibinfo{person}{Weibo Gao}, \bibinfo{person}{Qi Liu}, \bibinfo{person}{Zhenya Huang}, \bibinfo{person}{Yu Yin}, \bibinfo{person}{Haoyang Bi}, \bibinfo{person}{Mu-Chun Wang}, \bibinfo{person}{Jianhui Ma}, \bibinfo{person}{Shijin Wang}, {and} \bibinfo{person}{Yu Su}.} \bibinfo{year}{2021}\natexlab{}.
\newblock \showarticletitle{Rcd: Relation map driven cognitive diagnosis for intelligent education systems}. In \bibinfo{booktitle}{\emph{Proceedings of the 44th International ACM SIGIR Conference on Research and Development in Information Retrieval}}. \bibinfo{pages}{501--510}.
\newblock


\bibitem[Gao et~al\mbox{.}(2024)]%
        {gao2024zero}
\bibfield{author}{\bibinfo{person}{Weibo Gao}, \bibinfo{person}{Qi Liu}, \bibinfo{person}{Hao Wang}, \bibinfo{person}{Linan Yue}, \bibinfo{person}{Haoyang Bi}, \bibinfo{person}{Yin Gu}, \bibinfo{person}{Fangzhou Yao}, \bibinfo{person}{Zheng Zhang}, \bibinfo{person}{Xin Li}, {and} \bibinfo{person}{Yuanjing He}.} \bibinfo{year}{2024}\natexlab{}.
\newblock \showarticletitle{Zero-1-to-3: Domain-Level Zero-Shot Cognitive Diagnosis via One Batch of Early-Bird Students towards Three Diagnostic Objectives}. In \bibinfo{booktitle}{\emph{Proceedings of the AAAI Conference on Artificial Intelligence}}, Vol.~\bibinfo{volume}{38}. \bibinfo{pages}{8417--8426}.
\newblock


\bibitem[Ghosh et~al\mbox{.}(2020)]%
        {ghosh2020context}
\bibfield{author}{\bibinfo{person}{Aritra Ghosh}, \bibinfo{person}{Neil Heffernan}, {and} \bibinfo{person}{Andrew~S Lan}.} \bibinfo{year}{2020}\natexlab{}.
\newblock \showarticletitle{Context-aware attentive knowledge tracing}. In \bibinfo{booktitle}{\emph{Proceedings of the 26th ACM SIGKDD international conference on knowledge discovery \& data mining}}. \bibinfo{pages}{2330--2339}.
\newblock


\bibitem[Glorot and Bengio(2010)]%
        {glorot2010understanding}
\bibfield{author}{\bibinfo{person}{Xavier Glorot} {and} \bibinfo{person}{Yoshua Bengio}.} \bibinfo{year}{2010}\natexlab{}.
\newblock \showarticletitle{Understanding the difficulty of training deep feedforward neural networks}. In \bibinfo{booktitle}{\emph{Proceedings of the thirteenth international conference on artificial intelligence and statistics}}. JMLR Workshop and Conference Proceedings, \bibinfo{pages}{249--256}.
\newblock


\bibitem[Han et~al\mbox{.}(2023)]%
        {han2023errorclr}
\bibfield{author}{\bibinfo{person}{Siqi Han}, \bibinfo{person}{Yu Wang}, {and} \bibinfo{person}{Xuesong Lu}.} \bibinfo{year}{2023}\natexlab{}.
\newblock \showarticletitle{ErrorCLR: Semantic Error Classification, Localization and Repair for Introductory Programming Assignments}. In \bibinfo{booktitle}{\emph{Proceedings of the 46th International ACM SIGIR Conference on Research and Development in Information Retrieval}}. \bibinfo{pages}{1345--1354}.
\newblock


\bibitem[Hartigan and Wong(1979)]%
        {hartigan1979algorithm}
\bibfield{author}{\bibinfo{person}{John~A Hartigan} {and} \bibinfo{person}{Manchek~A Wong}.} \bibinfo{year}{1979}\natexlab{}.
\newblock \showarticletitle{Algorithm AS 136: A k-means clustering algorithm}.
\newblock \bibinfo{journal}{\emph{Journal of the royal statistical society. series c (applied statistics)}} \bibinfo{volume}{28}, \bibinfo{number}{1} (\bibinfo{year}{1979}), \bibinfo{pages}{100--108}.
\newblock


\bibitem[Ho et~al\mbox{.}(2020)]%
        {ho2020denoising}
\bibfield{author}{\bibinfo{person}{Jonathan Ho}, \bibinfo{person}{Ajay Jain}, {and} \bibinfo{person}{Pieter Abbeel}.} \bibinfo{year}{2020}\natexlab{}.
\newblock \showarticletitle{Denoising diffusion probabilistic models}.
\newblock \bibinfo{journal}{\emph{Advances in neural information processing systems}}  \bibinfo{volume}{33} (\bibinfo{year}{2020}), \bibinfo{pages}{6840--6851}.
\newblock


\bibitem[Hochreiter and Schmidhuber(1997)]%
        {hochreiter1997long}
\bibfield{author}{\bibinfo{person}{Sepp Hochreiter} {and} \bibinfo{person}{J{\"u}rgen Schmidhuber}.} \bibinfo{year}{1997}\natexlab{}.
\newblock \showarticletitle{Long short-term memory}.
\newblock \bibinfo{journal}{\emph{Neural computation}} \bibinfo{volume}{9}, \bibinfo{number}{8} (\bibinfo{year}{1997}), \bibinfo{pages}{1735--1780}.
\newblock


\bibitem[Hu et~al\mbox{.}(2021)]%
        {hu2021lora}
\bibfield{author}{\bibinfo{person}{Edward~J Hu}, \bibinfo{person}{Yelong Shen}, \bibinfo{person}{Phillip Wallis}, \bibinfo{person}{Zeyuan Allen-Zhu}, \bibinfo{person}{Yuanzhi Li}, \bibinfo{person}{Shean Wang}, \bibinfo{person}{Lu Wang}, {and} \bibinfo{person}{Weizhu Chen}.} \bibinfo{year}{2021}\natexlab{}.
\newblock \showarticletitle{Lora: Low-rank adaptation of large language models}.
\newblock \bibinfo{journal}{\emph{arXiv preprint arXiv:2106.09685}} (\bibinfo{year}{2021}).
\newblock


\bibitem[Jin et~al\mbox{.}(2023)]%
        {jin2023transferable}
\bibfield{author}{\bibinfo{person}{Yilun Jin}, \bibinfo{person}{Kai Chen}, {and} \bibinfo{person}{Qiang Yang}.} \bibinfo{year}{2023}\natexlab{}.
\newblock \showarticletitle{Transferable graph structure learning for graph-based traffic forecasting across cities}. In \bibinfo{booktitle}{\emph{Proceedings of the 29th ACM SIGKDD Conference on Knowledge Discovery and Data Mining}}. \bibinfo{pages}{1032--1043}.
\newblock


\bibitem[Kim et~al\mbox{.}(2016)]%
        {kim2016hadamard}
\bibfield{author}{\bibinfo{person}{Jin-Hwa Kim}, \bibinfo{person}{Kyoung-Woon On}, \bibinfo{person}{Woosang Lim}, \bibinfo{person}{Jeonghee Kim}, \bibinfo{person}{Jung-Woo Ha}, {and} \bibinfo{person}{Byoung-Tak Zhang}.} \bibinfo{year}{2016}\natexlab{}.
\newblock \showarticletitle{Hadamard product for low-rank bilinear pooling}.
\newblock \bibinfo{journal}{\emph{arXiv preprint arXiv:1610.04325}} (\bibinfo{year}{2016}).
\newblock


\bibitem[Kingma and Ba(2014)]%
        {kingma2014adam}
\bibfield{author}{\bibinfo{person}{Diederik~P Kingma} {and} \bibinfo{person}{Jimmy Ba}.} \bibinfo{year}{2014}\natexlab{}.
\newblock \showarticletitle{Adam: A method for stochastic optimization}.
\newblock \bibinfo{journal}{\emph{arXiv preprint arXiv:1412.6980}} (\bibinfo{year}{2014}).
\newblock


\bibitem[Lei and Mendes(2021)]%
        {lei2021systematic}
\bibfield{author}{\bibinfo{person}{Philip~IS Lei} {and} \bibinfo{person}{Ant{\'o}nio~Jos{\'e} Mendes}.} \bibinfo{year}{2021}\natexlab{}.
\newblock \showarticletitle{A systematic literature review on knowledge tracing in learning programming}. In \bibinfo{booktitle}{\emph{2021 IEEE Frontiers in Education Conference (FIE)}}. IEEE, \bibinfo{pages}{1--7}.
\newblock


\bibitem[Li et~al\mbox{.}(2021)]%
        {li2021intention}
\bibfield{author}{\bibinfo{person}{Haoyang Li}, \bibinfo{person}{Xin Wang}, \bibinfo{person}{Ziwei Zhang}, \bibinfo{person}{Jianxin Ma}, \bibinfo{person}{Peng Cui}, {and} \bibinfo{person}{Wenwu Zhu}.} \bibinfo{year}{2021}\natexlab{}.
\newblock \showarticletitle{Intention-aware sequential recommendation with structured intent transition}.
\newblock \bibinfo{journal}{\emph{IEEE Transactions on Knowledge and Data Engineering}} \bibinfo{volume}{34}, \bibinfo{number}{11} (\bibinfo{year}{2021}), \bibinfo{pages}{5403--5414}.
\newblock


\bibitem[Li et~al\mbox{.}(2025)]%
        {li2025foundation}
\bibfield{author}{\bibinfo{person}{Mingjia Li}, \bibinfo{person}{Hong Qian}, \bibinfo{person}{Jinglan Lv}, \bibinfo{person}{Mengliang He}, \bibinfo{person}{Wei Zhang}, {and} \bibinfo{person}{Aimin Zhou}.} \bibinfo{year}{2025}\natexlab{}.
\newblock \showarticletitle{Foundation model enhanced derivative-free cognitive diagnosis}.
\newblock \bibinfo{journal}{\emph{Frontiers of Computer Science}} \bibinfo{volume}{19}, \bibinfo{number}{1} (\bibinfo{year}{2025}), \bibinfo{pages}{191318}.
\newblock


\bibitem[Li et~al\mbox{.}(2022a)]%
        {li2022pst}
\bibfield{author}{\bibinfo{person}{Ruixin Li}, \bibinfo{person}{Yu Yin}, \bibinfo{person}{Le Dai}, \bibinfo{person}{Shuanghong Shen}, \bibinfo{person}{Xin Lin}, \bibinfo{person}{Yu Su}, {and} \bibinfo{person}{Enhong Chen}.} \bibinfo{year}{2022}\natexlab{a}.
\newblock \showarticletitle{PST: Measuring Skill Proficiency in Programming Exercise Process via Programming Skill Tracing}. In \bibinfo{booktitle}{\emph{Proceedings of the 45th International ACM SIGIR Conference on Research and Development in Information Retrieval}}. \bibinfo{pages}{2601--2606}.
\newblock


\bibitem[Li et~al\mbox{.}(2022b)]%
        {li2022mining}
\bibfield{author}{\bibinfo{person}{Rongfan Li}, \bibinfo{person}{Ting Zhong}, \bibinfo{person}{Xinke Jiang}, \bibinfo{person}{Goce Trajcevski}, \bibinfo{person}{Jin Wu}, {and} \bibinfo{person}{Fan Zhou}.} \bibinfo{year}{2022}\natexlab{b}.
\newblock \showarticletitle{Mining spatio-temporal relations via self-paced graph contrastive learning}. In \bibinfo{booktitle}{\emph{Proceedings of the 28th ACM SIGKDD Conference on Knowledge Discovery and Data Mining}}. \bibinfo{pages}{936--944}.
\newblock


\bibitem[Liang et~al\mbox{.}(2022)]%
        {liang2022help}
\bibfield{author}{\bibinfo{person}{Yu Liang}, \bibinfo{person}{Tianhao Peng}, \bibinfo{person}{Yanjun Pu}, {and} \bibinfo{person}{Wenjun Wu}.} \bibinfo{year}{2022}\natexlab{}.
\newblock \showarticletitle{HELP-DKT: an interpretable cognitive model of how students learn programming based on deep knowledge tracing}.
\newblock \bibinfo{journal}{\emph{Scientific Reports}} \bibinfo{volume}{12}, \bibinfo{number}{1} (\bibinfo{year}{2022}), \bibinfo{pages}{1--11}.
\newblock


\bibitem[Liu et~al\mbox{.}(2023)]%
        {liu2023guiding}
\bibfield{author}{\bibinfo{person}{Jiayu Liu}, \bibinfo{person}{Zhenya Huang}, \bibinfo{person}{Zhiyuan Ma}, \bibinfo{person}{Qi Liu}, \bibinfo{person}{Enhong Chen}, \bibinfo{person}{Tianhuang Su}, {and} \bibinfo{person}{Haifeng Liu}.} \bibinfo{year}{2023}\natexlab{}.
\newblock \showarticletitle{Guiding Mathematical Reasoning via Mastering Commonsense Formula Knowledge}. In \bibinfo{booktitle}{\emph{Proceedings of the 29th ACM SIGKDD Conference on Knowledge Discovery and Data Mining}}. \bibinfo{pages}{1477--1488}.
\newblock


\bibitem[Liu et~al\mbox{.}(2022)]%
        {liu2022gpt}
\bibfield{author}{\bibinfo{person}{Naiming Liu}, \bibinfo{person}{Zichao Wang}, \bibinfo{person}{Richard~G Baraniuk}, {and} \bibinfo{person}{Andrew Lan}.} \bibinfo{year}{2022}\natexlab{}.
\newblock \showarticletitle{GPT-based Open-Ended Knowledge Tracing}.
\newblock \bibinfo{journal}{\emph{arXiv preprint arXiv:2203.03716}} (\bibinfo{year}{2022}).
\newblock


\bibitem[Liu et~al\mbox{.}(2019)]%
        {liu2019ekt}
\bibfield{author}{\bibinfo{person}{Qi Liu}, \bibinfo{person}{Zhenya Huang}, \bibinfo{person}{Yu Yin}, \bibinfo{person}{Enhong Chen}, \bibinfo{person}{Hui Xiong}, \bibinfo{person}{Yu Su}, {and} \bibinfo{person}{Guoping Hu}.} \bibinfo{year}{2019}\natexlab{}.
\newblock \showarticletitle{Ekt: Exercise-aware knowledge tracing for student performance prediction}.
\newblock \bibinfo{journal}{\emph{IEEE Transactions on Knowledge and Data Engineering}} \bibinfo{volume}{33}, \bibinfo{number}{1} (\bibinfo{year}{2019}), \bibinfo{pages}{100--115}.
\newblock


\bibitem[Liu et~al\mbox{.}(2024)]%
        {Liu2024Icdm}
\bibfield{author}{\bibinfo{person}{Shuo Liu}, \bibinfo{person}{Junhao Shen}, \bibinfo{person}{Hong Qian}, {and} \bibinfo{person}{Aimin Zhou}.} \bibinfo{year}{2024}\natexlab{}.
\newblock \showarticletitle{Inductive Cognitive Diagnosis for Fast Student Learning in Web-Based Intelligent Education Systems}. In \bibinfo{booktitle}{\emph{Proceedings of the {ACM} on Web Conference 2024}}. \bibinfo{address}{Singapore}, \bibinfo{pages}{4260--4271}.
\newblock


\bibitem[Ma et~al\mbox{.}(2024)]%
        {ma2024hd}
\bibfield{author}{\bibinfo{person}{Haiping Ma}, \bibinfo{person}{Yong Yang}, \bibinfo{person}{Chuan Qin}, \bibinfo{person}{Xiaoshan Yu}, \bibinfo{person}{Shangshang Yang}, \bibinfo{person}{Xingyi Zhang}, {and} \bibinfo{person}{Hengshu Zhu}.} \bibinfo{year}{2024}\natexlab{}.
\newblock \showarticletitle{HD-KT: Advancing Robust Knowledge Tracing via Anomalous Learning Interaction Detection}. In \bibinfo{booktitle}{\emph{Proceedings of the ACM on Web Conference 2024}}. \bibinfo{pages}{4479--4488}.
\newblock


\bibitem[Nguyen et~al\mbox{.}(2014)]%
        {nguyen2014codewebs}
\bibfield{author}{\bibinfo{person}{Andy Nguyen}, \bibinfo{person}{Christopher Piech}, \bibinfo{person}{Jonathan Huang}, {and} \bibinfo{person}{Leonidas Guibas}.} \bibinfo{year}{2014}\natexlab{}.
\newblock \showarticletitle{Codewebs: scalable homework search for massive open online programming courses}. In \bibinfo{booktitle}{\emph{Proceedings of the 23rd international conference on World wide web}}. \bibinfo{pages}{491--502}.
\newblock


\bibitem[Paszke et~al\mbox{.}(2019)]%
        {paszke2019pytorch}
\bibfield{author}{\bibinfo{person}{Adam Paszke}, \bibinfo{person}{Sam Gross}, \bibinfo{person}{Francisco Massa}, \bibinfo{person}{Adam Lerer}, \bibinfo{person}{James Bradbury}, \bibinfo{person}{Gregory Chanan}, \bibinfo{person}{Trevor Killeen}, \bibinfo{person}{Zeming Lin}, \bibinfo{person}{Natalia Gimelshein}, \bibinfo{person}{Luca Antiga}, {et~al\mbox{.}}} \bibinfo{year}{2019}\natexlab{}.
\newblock \showarticletitle{Pytorch: An imperative style, high-performance deep learning library}.
\newblock \bibinfo{journal}{\emph{Advances in neural information processing systems}}  \bibinfo{volume}{32} (\bibinfo{year}{2019}).
\newblock


\bibitem[Piech et~al\mbox{.}(2015)]%
        {piech2015deep}
\bibfield{author}{\bibinfo{person}{Chris Piech}, \bibinfo{person}{Jonathan Bassen}, \bibinfo{person}{Jonathan Huang}, \bibinfo{person}{Surya Ganguli}, \bibinfo{person}{Mehran Sahami}, \bibinfo{person}{Leonidas~J Guibas}, {and} \bibinfo{person}{Jascha Sohl-Dickstein}.} \bibinfo{year}{2015}\natexlab{}.
\newblock \showarticletitle{Deep knowledge tracing}.
\newblock \bibinfo{journal}{\emph{Advances in neural information processing systems}}  \bibinfo{volume}{28} (\bibinfo{year}{2015}).
\newblock


\bibitem[Puri et~al\mbox{.}(2021)]%
        {puri2021codenet}
\bibfield{author}{\bibinfo{person}{Ruchir Puri}, \bibinfo{person}{David~S. Kung}, \bibinfo{person}{Geert Janssen}, \bibinfo{person}{Wei Zhang}, \bibinfo{person}{Giacomo Domeniconi}, \bibinfo{person}{Vladimir Zolotov}, \bibinfo{person}{Julian Dolby}, \bibinfo{person}{Jie Chen}, \bibinfo{person}{Mihir Choudhury}, \bibinfo{person}{Lindsey Decker}, \bibinfo{person}{Veronika Thost}, \bibinfo{person}{Luca Buratti}, \bibinfo{person}{Saurabh Pujar}, \bibinfo{person}{Shyam Ramji}, \bibinfo{person}{Ulrich Finkler}, \bibinfo{person}{Susan Malaika}, {and} \bibinfo{person}{Frederick Reiss}.} \bibinfo{year}{2021}\natexlab{}.
\newblock \bibinfo{title}{CodeNet: A Large-Scale AI for Code Dataset for Learning a Diversity of Coding Tasks}.
\newblock
\showeprint[arxiv]{2105.12655}~[cs.SE]


\bibitem[Ruder(2016)]%
        {ruder2016overview}
\bibfield{author}{\bibinfo{person}{Sebastian Ruder}.} \bibinfo{year}{2016}\natexlab{}.
\newblock \showarticletitle{An overview of gradient descent optimization algorithms}.
\newblock \bibinfo{journal}{\emph{arXiv preprint arXiv:1609.04747}} (\bibinfo{year}{2016}).
\newblock


\bibitem[Shen et~al\mbox{.}(2021)]%
        {ShenS021lpckt}
\bibfield{author}{\bibinfo{person}{Shuanghong Shen}, \bibinfo{person}{Qi Liu}, \bibinfo{person}{Enhong Chen}, \bibinfo{person}{Zhenya Huang}, \bibinfo{person}{Wei Huang}, \bibinfo{person}{Yu Yin}, \bibinfo{person}{Yu Su}, {and} \bibinfo{person}{Shijin Wang}.} \bibinfo{year}{2021}\natexlab{}.
\newblock \showarticletitle{Learning Process-consistent Knowledge Tracing}. In \bibinfo{booktitle}{\emph{{KDD} '21: The 27th {ACM} {SIGKDD} Conference on Knowledge Discovery and Data Mining, Virtual Event, Singapore, August 14-18, 2021}}, \bibfield{editor}{\bibinfo{person}{Feida Zhu}, \bibinfo{person}{Beng~Chin Ooi}, {and} \bibinfo{person}{Chunyan Miao}} (Eds.). \bibinfo{publisher}{{ACM}}, \bibinfo{pages}{1452--1460}.
\newblock
\href{https://doi.org/10.1145/3447548.3467237}{doi:\nolinkurl{10.1145/3447548.3467237}}


\bibitem[Shen et~al\mbox{.}(2024)]%
        {shen2024survey}
\bibfield{author}{\bibinfo{person}{Shuanghong Shen}, \bibinfo{person}{Qi Liu}, \bibinfo{person}{Zhenya Huang}, \bibinfo{person}{Yonghe Zheng}, \bibinfo{person}{Minghao Yin}, \bibinfo{person}{Minjuan Wang}, {and} \bibinfo{person}{Enhong Chen}.} \bibinfo{year}{2024}\natexlab{}.
\newblock \showarticletitle{A survey of knowledge tracing: Models, variants, and applications}.
\newblock \bibinfo{journal}{\emph{IEEE Transactions on Learning Technologies}} (\bibinfo{year}{2024}).
\newblock


\bibitem[Shi et~al\mbox{.}(2022)]%
        {shi2022code}
\bibfield{author}{\bibinfo{person}{Yang Shi}, \bibinfo{person}{Min Chi}, \bibinfo{person}{Tiffany Barnes}, {and} \bibinfo{person}{Thomas Price}.} \bibinfo{year}{2022}\natexlab{}.
\newblock \showarticletitle{Code-DKT: A Code-based Knowledge Tracing Model for Programming Tasks}.
\newblock \bibinfo{journal}{\emph{arXiv preprint arXiv:2206.03545}} (\bibinfo{year}{2022}).
\newblock


\bibitem[Wang et~al\mbox{.}(2024)]%
        {wang2024survey}
\bibfield{author}{\bibinfo{person}{Fei Wang}, \bibinfo{person}{Weibo Gao}, \bibinfo{person}{Qi Liu}, \bibinfo{person}{Jiatong Li}, \bibinfo{person}{Guanhao Zhao}, \bibinfo{person}{Zheng Zhang}, \bibinfo{person}{Zhenya Huang}, \bibinfo{person}{Mengxiao Zhu}, \bibinfo{person}{Shijin Wang}, \bibinfo{person}{Wei Tong}, {et~al\mbox{.}}} \bibinfo{year}{2024}\natexlab{}.
\newblock \showarticletitle{A survey of models for cognitive diagnosis: New developments and future directions}.
\newblock \bibinfo{journal}{\emph{arXiv preprint arXiv:2407.05458}} (\bibinfo{year}{2024}).
\newblock


\bibitem[Wang et~al\mbox{.}(2022)]%
        {wang2022neuralcd}
\bibfield{author}{\bibinfo{person}{Fei Wang}, \bibinfo{person}{Qi Liu}, \bibinfo{person}{Enhong Chen}, \bibinfo{person}{Zhenya Huang}, \bibinfo{person}{Yu Yin}, \bibinfo{person}{Shijin Wang}, {and} \bibinfo{person}{Yu Su}.} \bibinfo{year}{2022}\natexlab{}.
\newblock \showarticletitle{NeuralCD: a general framework for cognitive diagnosis}.
\newblock \bibinfo{journal}{\emph{IEEE Transactions on Knowledge and Data Engineering}} \bibinfo{volume}{35}, \bibinfo{number}{8} (\bibinfo{year}{2022}), \bibinfo{pages}{8312--8327}.
\newblock


\bibitem[Wang et~al\mbox{.}(2017)]%
        {wang2017deep}
\bibfield{author}{\bibinfo{person}{Lisa Wang}, \bibinfo{person}{Angela Sy}, \bibinfo{person}{Larry Liu}, {and} \bibinfo{person}{Chris Piech}.} \bibinfo{year}{2017}\natexlab{}.
\newblock \showarticletitle{Deep knowledge tracing on programming exercises}. In \bibinfo{booktitle}{\emph{Proceedings of the fourth (2017) ACM conference on learning@ scale}}. \bibinfo{pages}{201--204}.
\newblock


\bibitem[Wang et~al\mbox{.}(2020)]%
        {wang2020gcn}
\bibfield{author}{\bibinfo{person}{Xiao Wang}, \bibinfo{person}{Meiqi Zhu}, \bibinfo{person}{Deyu Bo}, \bibinfo{person}{Peng Cui}, \bibinfo{person}{Chuan Shi}, {and} \bibinfo{person}{Jian Pei}.} \bibinfo{year}{2020}\natexlab{}.
\newblock \showarticletitle{Am-gcn: Adaptive multi-channel graph convolutional networks}. In \bibinfo{booktitle}{\emph{Proceedings of the 26th ACM SIGKDD International conference on knowledge discovery \& data mining}}. \bibinfo{pages}{1243--1253}.
\newblock


\bibitem[Wilamowski and Yu(2010)]%
        {wilamowski2010improved}
\bibfield{author}{\bibinfo{person}{Bogdan~M Wilamowski} {and} \bibinfo{person}{Hao Yu}.} \bibinfo{year}{2010}\natexlab{}.
\newblock \showarticletitle{Improved computation for Levenberg--Marquardt training}.
\newblock \bibinfo{journal}{\emph{IEEE transactions on neural networks}} \bibinfo{volume}{21}, \bibinfo{number}{6} (\bibinfo{year}{2010}), \bibinfo{pages}{930--937}.
\newblock


\bibitem[Wu et~al\mbox{.}(2024)]%
        {wu2024programming}
\bibfield{author}{\bibinfo{person}{Yaqiang Wu}, \bibinfo{person}{Hui Zhu}, \bibinfo{person}{Chenyang Wang}, \bibinfo{person}{Fujian Song}, \bibinfo{person}{Haiping Zhu}, \bibinfo{person}{Yan Chen}, \bibinfo{person}{Qinghua Zheng}, {and} \bibinfo{person}{Feng Tian}.} \bibinfo{year}{2024}\natexlab{}.
\newblock \showarticletitle{Programming Knowledge Tracing Based on Heterogeneous Graph Representation}.
\newblock \bibinfo{journal}{\emph{Knowledge-Based Systems}} (\bibinfo{year}{2024}), \bibinfo{pages}{112161}.
\newblock


\bibitem[Wu et~al\mbox{.}(2020)]%
        {wu2020exercise}
\bibfield{author}{\bibinfo{person}{Zhengyang Wu}, \bibinfo{person}{Ming Li}, \bibinfo{person}{Yong Tang}, {and} \bibinfo{person}{Qingyu Liang}.} \bibinfo{year}{2020}\natexlab{}.
\newblock \showarticletitle{Exercise recommendation based on knowledge concept prediction}.
\newblock \bibinfo{journal}{\emph{Knowledge-Based Systems}}  \bibinfo{volume}{210} (\bibinfo{year}{2020}), \bibinfo{pages}{106481}.
\newblock


\bibitem[Xia et~al\mbox{.}(2021)]%
        {xia2021graph}
\bibfield{author}{\bibinfo{person}{Feng Xia}, \bibinfo{person}{Ke Sun}, \bibinfo{person}{Shuo Yu}, \bibinfo{person}{Abdul Aziz}, \bibinfo{person}{Liangtian Wan}, \bibinfo{person}{Shirui Pan}, {and} \bibinfo{person}{Huan Liu}.} \bibinfo{year}{2021}\natexlab{}.
\newblock \showarticletitle{Graph learning: A survey}.
\newblock \bibinfo{journal}{\emph{IEEE Transactions on Artificial Intelligence}} \bibinfo{volume}{2}, \bibinfo{number}{2} (\bibinfo{year}{2021}), \bibinfo{pages}{109--127}.
\newblock


\bibitem[Xu and Wunsch(2005)]%
        {xu2005survey}
\bibfield{author}{\bibinfo{person}{Rui Xu} {and} \bibinfo{person}{Donald Wunsch}.} \bibinfo{year}{2005}\natexlab{}.
\newblock \showarticletitle{Survey of clustering algorithms}.
\newblock \bibinfo{journal}{\emph{IEEE Transactions on neural networks}} \bibinfo{volume}{16}, \bibinfo{number}{3} (\bibinfo{year}{2005}), \bibinfo{pages}{645--678}.
\newblock


\bibitem[Yao et~al\mbox{.}(2024)]%
        {yao2024adard}
\bibfield{author}{\bibinfo{person}{Fangzhou Yao}, \bibinfo{person}{Qi Liu}, \bibinfo{person}{Linan Yue}, \bibinfo{person}{Weibo Gao}, \bibinfo{person}{Jiatong Li}, \bibinfo{person}{Xin Li}, {and} \bibinfo{person}{Yuanjing He}.} \bibinfo{year}{2024}\natexlab{}.
\newblock \showarticletitle{Adard: An adaptive response denoising framework for robust learner modeling}. In \bibinfo{booktitle}{\emph{Proceedings of the 30th ACM SIGKDD Conference on Knowledge Discovery and Data Mining}}. \bibinfo{pages}{3886--3895}.
\newblock


\bibitem[Yin et~al\mbox{.}(2023)]%
        {yin2023tracing}
\bibfield{author}{\bibinfo{person}{Yu Yin}, \bibinfo{person}{Le Dai}, \bibinfo{person}{Zhenya Huang}, \bibinfo{person}{Shuanghong Shen}, \bibinfo{person}{Fei Wang}, \bibinfo{person}{Qi Liu}, \bibinfo{person}{Enhong Chen}, {and} \bibinfo{person}{Xin Li}.} \bibinfo{year}{2023}\natexlab{}.
\newblock \showarticletitle{Tracing Knowledge Instead of Patterns: Stable Knowledge Tracing with Diagnostic Transformer}. In \bibinfo{booktitle}{\emph{Proceedings of the ACM Web Conference 2023}}. \bibinfo{pages}{855--864}.
\newblock


\bibitem[Yu et~al\mbox{.}(2024)]%
        {yu2024eckt}
\bibfield{author}{\bibinfo{person}{Yang Yu}, \bibinfo{person}{Yingbo Zhou}, \bibinfo{person}{Yaokang Zhu}, \bibinfo{person}{Yutong Ye}, \bibinfo{person}{Liangyu Chen}, {and} \bibinfo{person}{Mingsong Chen}.} \bibinfo{year}{2024}\natexlab{}.
\newblock \showarticletitle{ECKT: Enhancing Code Knowledge Tracing via Large Language Models}. In \bibinfo{booktitle}{\emph{Proceedings of the Annual Meeting of the Cognitive Science Society}}, Vol.~\bibinfo{volume}{46}.
\newblock


\bibitem[Yue et~al\mbox{.}(2024)]%
        {yue2024cooperative}
\bibfield{author}{\bibinfo{person}{Linan Yue}, \bibinfo{person}{Qi Liu}, \bibinfo{person}{Ye Liu}, \bibinfo{person}{Weibo Gao}, \bibinfo{person}{Fangzhou Yao}, {and} \bibinfo{person}{Wenfeng Li}.} \bibinfo{year}{2024}\natexlab{}.
\newblock \showarticletitle{Cooperative classification and rationalization for graph generalization}. In \bibinfo{booktitle}{\emph{Proceedings of the ACM on Web Conference 2024}}. \bibinfo{pages}{344--352}.
\newblock


\bibitem[Zhang et~al\mbox{.}(2024)]%
        {zhang2024understanding}
\bibfield{author}{\bibinfo{person}{Zheng Zhang}, \bibinfo{person}{Le Wu}, \bibinfo{person}{Qi Liu}, \bibinfo{person}{Jiayu Liu}, \bibinfo{person}{Zhenya Huang}, \bibinfo{person}{Yu Yin}, \bibinfo{person}{Yan Zhuang}, \bibinfo{person}{Weibo Gao}, {and} \bibinfo{person}{Enhong Chen}.} \bibinfo{year}{2024}\natexlab{}.
\newblock \showarticletitle{Understanding and improving fairness in cognitive diagnosis}.
\newblock \bibinfo{journal}{\emph{Science China Information Sciences}} \bibinfo{volume}{67}, \bibinfo{number}{5} (\bibinfo{year}{2024}), \bibinfo{pages}{152106}.
\newblock


\bibitem[Zhao et~al\mbox{.}(2023)]%
        {zhao2023cross}
\bibfield{author}{\bibinfo{person}{Chuang Zhao}, \bibinfo{person}{Hongke Zhao}, \bibinfo{person}{Xiaomeng Li}, \bibinfo{person}{Ming He}, \bibinfo{person}{Jiahui Wang}, {and} \bibinfo{person}{Jianping Fan}.} \bibinfo{year}{2023}\natexlab{}.
\newblock \showarticletitle{Cross-domain recommendation via progressive structural alignment}.
\newblock \bibinfo{journal}{\emph{IEEE Transactions on Knowledge and Data Engineering}} (\bibinfo{year}{2023}).
\newblock


\bibitem[Zhao et~al\mbox{.}(2017)]%
        {zhao2017physics}
\bibfield{author}{\bibinfo{person}{Hao Zhao}, \bibinfo{person}{Ming Lu}, \bibinfo{person}{Anbang Yao}, \bibinfo{person}{Yiwen Guo}, \bibinfo{person}{Yurong Chen}, {and} \bibinfo{person}{Li Zhang}.} \bibinfo{year}{2017}\natexlab{}.
\newblock \showarticletitle{Physics inspired optimization on semantic transfer features: An alternative method for room layout estimation}. In \bibinfo{booktitle}{\emph{Proceedings of the IEEE conference on computer vision and pattern recognition}}. \bibinfo{pages}{10--18}.
\newblock


\bibitem[Zhou et~al\mbox{.}(2021)]%
        {zhou2021image}
\bibfield{author}{\bibinfo{person}{Man Zhou}, \bibinfo{person}{Jie Xiao}, \bibinfo{person}{Yifan Chang}, \bibinfo{person}{Xueyang Fu}, \bibinfo{person}{Aiping Liu}, \bibinfo{person}{Jinshan Pan}, {and} \bibinfo{person}{Zheng-Jun Zha}.} \bibinfo{year}{2021}\natexlab{}.
\newblock \showarticletitle{Image de-raining via continual learning}. In \bibinfo{booktitle}{\emph{Proceedings of the IEEE/CVF Conference on Computer Vision and Pattern Recognition}}. \bibinfo{pages}{4907--4916}.
\newblock


\bibitem[Zhu et~al\mbox{.}(2022)]%
        {zhu2022programming}
\bibfield{author}{\bibinfo{person}{Renyu Zhu}, \bibinfo{person}{Dongxiang Zhang}, \bibinfo{person}{Chengcheng Han}, \bibinfo{person}{Ming Gaol}, \bibinfo{person}{Xuesong Lu}, \bibinfo{person}{Weining Qian}, {and} \bibinfo{person}{Aoying Zhou}.} \bibinfo{year}{2022}\natexlab{}.
\newblock \showarticletitle{Programming knowledge tracing: A comprehensive dataset and a new model}. In \bibinfo{booktitle}{\emph{2022 IEEE International Conference on Data Mining Workshops (ICDMW)}}. IEEE, \bibinfo{pages}{298--307}.
\newblock


\bibitem[Zinovieva et~al\mbox{.}(2021)]%
        {zinovieva2021use}
\bibfield{author}{\bibinfo{person}{IS Zinovieva}, \bibinfo{person}{VO Artemchuk}, \bibinfo{person}{Anna~V Iatsyshyn}, \bibinfo{person}{OO Popov}, \bibinfo{person}{VO Kovach}, \bibinfo{person}{Andrii~V Iatsyshyn}, \bibinfo{person}{YO Romanenko}, {and} \bibinfo{person}{OV Radchenko}.} \bibinfo{year}{2021}\natexlab{}.
\newblock \showarticletitle{The use of online coding platforms as additional distance tools in programming education}. In \bibinfo{booktitle}{\emph{Journal of physics: Conference series}}, Vol.~\bibinfo{volume}{1840}. IOP Publishing, \bibinfo{pages}{012029}.
\newblock


\end{thebibliography}

\appendix

\section{Proof}
\label{app:proof}

\textbf{Property 1.}
$\mathcal{L}^{nav}_b$ \textit{is bounded as follows:}
\begin{equation}
    \begin{split}
    \mathcal{L}^{nav}_b \leq \left|\left(\nabla_{\theta_{b-1}}\mathcal{L}^{pkt}(\theta_{b-1})\right)^{T}\right| \cdot\left|\delta \theta_{b-1}\right|+\left|\delta \theta_{b-1}\right|^T \cdot|J^{T}J| \cdot\left|\delta \theta_{b-1}\right|,
    \label{eq:app_l_nav_bound}
    \end{split}
\end{equation}
where $J$ is the Jacobian matrix~\cite{wilamowski2010improved}.
The proof is as follows:
\begin{proof}
Given the following equations,
\begin{equation}
    \begin{split}
        \mathcal{L}^{nav}_b=\left|\sum_{u=1}^{N}\sum_{t=1}^{T_{u}}\mathcal{L}_{u,t}^{pkt}(\theta_{b})-\sum_{u=1}^{N}\sum_{t=1}^{T_{u}}\mathcal{L}_{u,t}^{pkt}(\theta_{b-1})\right|,\;\delta\theta_{b-1}=\theta_{b}-\theta_{b-1},
    \end{split}
\end{equation}
and let $\mathcal{L}^{pkt}(\theta_{b})$ and $\mathcal{L}^{pkt}(\theta_{b-1})$ denote $\sum_{u=1}^{N}\sum_{t=1}^{T_{u}}\mathcal{L}_{u,t}^{pkt}(\theta_{b})$ and $\sum_{u=1}^{N}\sum_{t=1}^{T_{u}}\mathcal{L}_{u,t}^{pkt}(\theta_{b-1})$ respectively, we take the Taylor expansion of $\mathcal{L}^{pkt}(\theta_{b-1})$ at point $\theta_{b-1}$, which is an infinite sum of terms that are expressed in the form of loss function’s derivatives at a single point:
\begin{equation}
    \begin{split}
        \mathcal{L}^{pkt}(\theta_{b-1}+\delta\theta_{b-1})&=\mathcal{L}^{pkt}(\theta_{b-1})+\left(\nabla_{\theta_{b-1}}\mathcal{L}^{pkt}(\theta_{b-1})\right)^{T}\cdot \delta \theta_{b-1} \\
& +\frac{1}{2}\left(\delta \theta_{b-1}\right)^T \cdot H \cdot \delta \theta_{b-1}+O\left(\left\|\delta \theta_{b-1}\right\|^3\right),
    \end{split}
    \label{eq:app_original_l}
\end{equation}
where $H$ denotes Hessian matrix:
\begin{equation}
    \begin{split}
        H=\nabla_{\theta_{b-1}}\mathcal{L}^{pkt}(\theta_{b-1}).
    \end{split}
\end{equation}
Then $\mathcal{L}^{nav}_b$ can be derived as:
\begin{equation}
    \begin{split}
        \mathcal{L}^{nav}_b&=\left|\mathcal{L}^{pkt}(\theta_{b})-\mathcal{L}^{pkt}(\theta_{b-1})\right| \\
        &=\left|\mathcal{L}^{pkt}(\theta_{b-1}+\delta\theta_{b-1})-\mathcal{L}^{pkt}(\theta_{b-1})\right|\\
        &\approx\left|\left(\nabla_{\theta_{b-1}}\mathcal{L}^{pkt}(\theta_{b-1})\right)^{T}\cdot \delta \theta_{b-1}+\frac{1}{2}\left(\delta \theta_{b-1}\right)^T \cdot H \cdot \delta \theta_{b-1}\right|.
        \label{eq:app_performance}
    \end{split}
\end{equation}
It is evident that Eq.(\ref{eq:app_performance}) significantly reduces the computational demands on each training datum compared to Eq.(\ref{eq:app_original_l}), thereby accelerating the calculation of the total loss. However, the storage requirements associated with Hessian matrices remain substantial~\cite{zhou2021image}. Therefore, drawing inspiration from the Gauss-Newton method~\cite{wilamowski2010improved}, we approximate $H$ as $2\cdot J^{T}J$ for the Hessian matrix, where $J$ represents the Jacobian matrix. Consequently, we arrive at the final formulation of $\mathcal{L}^{nav}_b$ as follows:
\begin{equation}
    \begin{split}
        \mathcal{L}^{nav}_b&\approx\left|\left(\nabla_{\theta_{b-1}}\mathcal{L}^{pkt}(\theta_{b-1})\right)^{T}\cdot \delta \theta_{b-1}+\left(\delta \theta_{b-1}\right)^T \cdot J^{T}J \cdot \delta \theta_{b-1}\right|\\
        &\leq \left|\left(\nabla_{\theta_{b-1}}\mathcal{L}^{pkt}(\theta_{b-1})\right)^{T}\right| \cdot\left|\delta \theta_{b-1}\right|+\left|\delta \theta_{b-1}\right|^T \cdot|J^{T}J| \cdot\left|\delta \theta_{b-1}\right|\\
        \label{eq:app_final_l_nav}
    \end{split}
\end{equation}
where $|\cdot|$ in the Eq.~(\ref{eq:app_final_l_nav}) denotes element-wise absolute value in our implementation. The proof is completed.
\end{proof}

\vspace{-0.5cm}
\begin{algorithm}
\caption{The training and test of PKT backbone models}\label{train_test_algo}
\begin{algorithmic}[1]
\REQUIRE Training dataset $D_{train}$, test dataset $D_{test}$, learning rate $\alpha$, number of epochs $N$
\ENSURE Trained PKT backbone model $M$
\STATE Initialize model parameters $W$ randomly
\STATE \textbf{Training started}
\FOR{$epoch=1$ \TO $N$}
    \FORALL{$l_u \in D_{train}$}
        \STATE // $l_u$ denotes a programming sequence of learner $u$ and its length is $T_u$
        \FOR{$t=1$ \TO $T_u-1$}
            \STATE Compute programming performance at the next step $\hat{r}_{u,t+1} = M(l_{u,1:t}; W)$
            \STATE // $l_{u,1:t}$ is programming records of $u$ from step $1$ to $t$
            \STATE Compute cross-entropy loss $\mathcal{L}_{u,t}^{pkt} = BCE\left(r_{u,t+1},\hat{r}_{u,t+1}\right)$
            \STATE Update parameters $W = W - \alpha \nabla \mathcal{L}_{u,t}^{pkt}$
        \ENDFOR
    \ENDFOR
\ENDFOR
\STATE \textbf{Training completed}
\STATE \textbf{}
\STATE \textbf{Testing started}
\FORALL{$l_u \in D_{test}$}
    \FOR{$t=1$ \TO $T_u-1$}
        \STATE Compute programming performance at the next step $\hat{r}_{u,t+1} = M(l_u; W)$
        \STATE Compute performance metric using $r_{u,t+1}$ and $\hat{r}_{u,t+1}$
    \ENDFOR
\ENDFOR
\STATE \textbf{Testing completed}
\end{algorithmic}
\end{algorithm}

\vspace{-0.5cm}
\section{Datasets and Preprocessing}
\label{app:dataset}

\textbf{BePKT}\cite{zhu2022programming} comprises extensive programming submission records in several languages authored by college students. We retain submission records written in \textit{C}, \textit{C++}, \textit{Python}, and \textit{Java}.
\textbf{AtCoder} is collected from the online programming competition website \textit{Atcoder.org}, consisting of codes in multiple languages, as published by the CodeNet project\cite{puri2021codenet}. We retain submission records in \textit{C}, \textit{C++}, and \textit{Python} as the coding languages.
Furthermore, since PST~\cite{li2022pst} only supports single-language programming scenarios, we additionally sample submissions written in \textit{C++} from BePKT to generate the \textbf{BePKT\_C++} dataset and submissions written in \textit{C} from AtCoder to generate the \textbf{AtCoder\_C} dataset, enabling us to evaluate PST.
For each dataset, we exclude learners with fewer than five submissions to ensure that every learner has a sufficient number of records for tracing.
\begin{algorithm}[h]
\caption{The Algorithm of Coda (Part 1)}
\label{algo:coda-pkt}
\begin{algorithmic}[1]
\STATE \textbf{Input:} $N$ learners' programming sequences $\{l_1, l_2, \cdots, l_N\}$, the optimized backbone PKT model $\mathcal{M}_{PKT}$, correct solution feature set $\{\mathcal{S}_{q_j}\}_{j=1}^{M}$ of $M$ questions, training set $\mathcal{D}_{train}$, test set $\mathcal{D}_{test}$.
\STATE \textbf{Parameters ($\theta$):} $\mathbf{W}$, $\mathbf{W}_a$, $\mathbf{W}_c$, $\mathbf{W}_A$, $\mathbf{W}_B$.

\STATE \textbf{Function} \texttt{CodeGraphConstruction}($l_u$):
\STATE \quad // \textit{Construct code graph based on code semantic similarity.}
\STATE \quad // $l_{u}=\{(q_{u,1}, c_{u,1}, x_{u,1}, r_{u,1}), \dots, (q_{u,T_u}, c_{u,T_u}, x_{u,T_u}, r_{u,T_u})\}$
\STATE \quad \textbf{For each} pair of codes $<x_{u,i}, x_{u,j}>$ in $l_u$:
\STATE \quad \quad Calculate indicator matrix $M_{u,<i,j>}$.
\STATE \quad \textbf{End for}
\STATE \quad Generate binary adjacency matrix $A_{u}$ via $\text{Rank}_{\epsilon}$.
\STATE \quad \textit{// Note that $A_{u}$ may not guarantee a connected graph, and it may contain $n$ subgraphs.}
\STATE \quad \textbf{Return} code graph set: $\{G_{u,i}\}_{i=1}^{n}$, isolated code set: $\mathcal{I}_{u}$.

\STATE \textbf{Function} \texttt{UnwantedSignalsIdentification}($\mathcal{I}_{u}$, $\{\mathcal{S}_{q_j}\}_{j=1}^{M}$, $\{G_{u,i}\}_{i=1}^{n}$):
\STATE \quad // \textit{Identify unwanted submissions from isolated nodes.}
\STATE \quad Initialize an unwanted set $\mathcal{I}_{u}^{*} \leftarrow \emptyset$
\STATE \quad \textbf{For each} isolated node $i \in \mathcal{I}_{u}$:
\STATE \quad \quad Obtain question ID $q_{u,i}$ corresponding to code $i$
\STATE \quad \quad \textbf{For each} ${x}_{u,j'}$ in $\mathcal{S}_{q_{u,i}}$:
\STATE \quad \quad \quad $s_{i,j'} \leftarrow {\rm sim}(\mathbf{W} \odot \vec{x}_{u,i}, \mathbf{W} \odot \vec{x}_{u,j'})$.
\STATE \quad \quad \textbf{End for}
\STATE \quad \quad $S_{i} \leftarrow \{s_{i,j'}; j' \in \mathcal{S}_{q_{u,i}}\}$
\STATE \quad \quad \textbf{If} mean($S_i$) $<$ median($S_i$):
\STATE \quad \quad \quad Mark node $i$ as unwanted and $\mathcal{I}_{u}^{*} \cup \{i\}$.
\STATE \quad \quad \textbf{Else}:
\STATE \quad \quad \quad Link code node $i$ with $\{G_{u,i}\}_{i=1}^{n}$.
\STATE \quad \textbf{End for}
\STATE \quad \textbf{Return} unwanted set: $\mathcal{I}_{u}^{*}$, updated code graphs: $\{G_{u,i}\}_{i=1}^{n}$.

\STATE \textbf{Function} \texttt{ClusterAwareWeakSignalFusion}($\{G_{u,i}\}_{i=1}^{n}$, $\{\vec{x}_{u,i}\}_{i=1}^{T_u}$, $C_k$, $L$):
\STATE \quad // \textit{Apply} $L$-\textit{layer cluster-aware GCN to update node features.}
\STATE \quad $\{\vec{x'}_{u,i}\}_{i=1}^{T_u} \leftarrow \mathbf{GCN}(\{\vec{x}_{u,i}\}_{i=1}^{T_u}, \{G_{u,i}\}_{i=1}^{n}, \mathbf{W}, \mathbf{W_{a}}, \mathbf{W_{c}})$
\STATE \quad $C_k$ clusters $\leftarrow \text{KMeans}(\{\vec{x'}_{u,i}\}, C_k)$.
\STATE \quad Mark core submissions and weak signals for each cluster.
\STATE \quad \textbf{Return} a core submission and its related weak signal set for each cluster $cl$: $\{(C_{u,cl}, \mathcal{W}_{u,cl})\}_{cl=1}^{C_k}$.
\end{algorithmic}
\end{algorithm}

\vspace{-0.5cm}
\section{Algorithms}
\label{app:algo}
Algorithm~\ref{train_test_algo} shows the pseudo-code of training and testing a PKT backbone model.
Algorithms~\ref{algo:coda-pkt} and \ref{algo:coda-pkt2} show the workflow of Coda.

\begin{algorithm}
\caption{The Algorithm of Coda (Part 2)}
\label{algo:coda-pkt2}
\begin{algorithmic}[1]
\STATE \textbf{BEGIN MAIN FUNCTION:}
\STATE \textbf{Initialize:} learning rate $lr$, $epoch \leftarrow 0$, $BatchSize$, $TotalEpoch$, the number of clusters $C_k$, the number of GCN layers $L$.
\STATE // \textit{Training process:}
\STATE \textbf{repeat}
\STATE \quad \textbf{For} $b$ from $1$ to $|\mathcal{D}_{train}| / BatchSize$:
\STATE \quad \quad Initialize prediction loss $\mathcal{L}^{p} \leftarrow 0$
\STATE \quad \quad \textbf{For each} $u \in \mathcal{D}_{train_b}$:
\STATE \quad \quad \quad $\{G_{u,i}\}_{i=1}^{n}$, $\mathcal{I}_{u} \leftarrow$ \texttt{CodeGraphConstruction}($l_u$).
\STATE \quad \quad \quad $\mathcal{I}_{u}^{*}$, $\{G_{u,i}\}_{i=1}^{n} \leftarrow$ \texttt{UnwantedSignalsIdentification}($\mathcal{I}_{u}$, $\{\mathcal{S}_{q_j}\}_{j=1}^{M}$, $\{G_{u,i}\}_{i=1}^{n}$).
\STATE \quad \quad \quad $\{(C_{u,cl}, \mathcal{W}_{u,cl})\}_{cl=1}^{C_k} \leftarrow$ \texttt{ClusterAwareWeakSignalFusion}($\{G_{u,i}\}_{i=1}^{n}$, $\{\vec{x}_{u,i}\}_{i=1}^{T_u}$, $C_k$, $L$).

\STATE \quad \quad \quad // \textit{Tuning Adaptor for PKT}
\STATE \quad \quad \quad \textbf{For} $t$ from 1 to $T_u-1$:
\STATE \quad \quad \quad \quad Construct prompt $\vec{p}_u,t$ via $\mathcal{I}_{u}$, $\{(C_{u,cl}, \mathcal{W}_{u,cl})\}_{cl=1}^{C_k}$.
\STATE \quad \quad \quad \quad // \textit{Calculate correction signal.}
\STATE \quad \quad \quad \quad $\vec{h'}_{t_{u,a}} \leftarrow (\mathbf{W}_{A}^{T} \cdot \mathbf{W}_{B}) \odot \vec{p}_{u,t}$.
\STATE \quad \quad \quad \quad // \textit{Update knowledge state $\vec{h}_{u,t}$ learned by the PKT backbone $\mathcal{M}_{PKT}$.}
\STATE \quad \quad \quad \quad $\vec{h'}_{u,t} \leftarrow \vec{h}_{u,t} + \vec{h'}_{u,t_{a}}$.
\STATE \quad \quad \quad \quad // \textit{Compute programming performance at the step $t+1$ via programming records of $u$ from step $1$ to $t$.}
\STATE \quad \quad \quad \quad $\hat{r}_{u,t+1} = \mathcal{M}_{PKT}(l_{u,1:t}; \vec{h'}_{u,t})$
\STATE \quad \quad \quad \quad Compute loss $\mathcal{L}_{p} = \mathcal{L}_{u,t}^{pkt} + \mathcal{L}_{u,t}^{adaptor}$.
\STATE \quad \quad \quad \textbf{End for}
\STATE \quad \quad \textbf{End for}
\STATE \quad \quad Calculate $\theta_{b} \leftarrow \theta_{b-1} - lr \cdot \nabla_{\theta} \mathcal{L}^{p}$, without gradient backpropagating.

\STATE \quad \quad Calculate $\mathcal{L}^{nav}_b$ via $\theta_{b}$, $\theta_{b-1}$.
\STATE \quad \quad $\mathcal{L}^{Coda} \leftarrow \mathcal{L}_{p} + \mathcal{L}^{nav}_b$
\STATE \quad \quad Update parameters $\theta_{b} \leftarrow \theta_{b-1} - lr \cdot  \nabla_{\theta} \mathcal{L}^{nav}_b$, with gradient backpropagating.
\STATE \quad \textbf{End for}
\STATE \textbf{until} $epoch$ \textbf{equals} $TotalEpoch$

\STATE // \textit{Test process:}

\STATE \textbf{For each} $u \in \mathcal{D}_{test}$:
\STATE \quad \textbf{For} $t$ from 1 to $T_u-1$:
\STATE \quad \quad $\{G_{u,i}\}_{i=1}^{n}$, $\mathcal{I}_{u} \leftarrow$ \texttt{CodeGraphConstruction}($l_{u,1:t}$).
\STATE \quad \quad $\mathcal{I}_{u}^{*}$, $\{G_{u,i}\}_{i=1}^{n} \leftarrow$ \texttt{UnwantedSignalsIdentification}($\mathcal{I}_{u}$, $\{\mathcal{S}_{q_j}\}_{j=1}^{M}$, $\{G_{u,i}\}_{i=1}^{n}$).
\STATE \quad \quad $\{(C_{u,cl}, \mathcal{W}_{u,cl})\}_{cl=1}^{C_k} \leftarrow$ \texttt{ClusterAwareWeakSignalFusion}($\{G_{u,i}\}_{i=1}^{n}$, $\{\vec{x}_{u,i}\}_{i=1}^{T_u}$, $C_k$, $L$).
\STATE \quad \quad Construct prompt $\vec{p}_u,t$ via $\mathcal{I}_{u}$, $\{(C_{u,cl}, \mathcal{W}_{u,cl})\}_{cl=1}^{C_k}$.
\STATE \quad \quad $\vec{h'}_{t_{u,a}} \leftarrow (\mathbf{W}_{A}^{T} \cdot \mathbf{W}_{B}) \odot \vec{p}_{u,t}$.
\STATE \quad \quad $\vec{h'}_{u,t} \leftarrow \vec{h}_{u,t} + \vec{h'}_{u,t_{a}}$.
\STATE \quad \quad $\hat{r}_{u,t+1} = \mathcal{M}_{PKT}(l_{u,1:t}; \vec{h'}_{u,t})$
\STATE \quad \quad Compute performance metric using $r_{u,t+1}$ and $\hat{r}_{u,t+1}$.
\STATE \quad \textbf{End for}
\STATE \textbf{End for}
\end{algorithmic}
\end{algorithm}

\end{document}